\documentclass[twocolumn,apj,numberedappendix]{openjournal}

\usepackage{lipsum}

\usepackage{xcolor}
\usepackage{textgreek}
\usepackage{enumitem}
\usepackage[utf8]{inputenc}
\usepackage[english]{babel}

\usepackage{hyperref}
\hypersetup{
    unicode, 
    colorlinks=true,
    linkcolor=linkcolor,
    citecolor=linkcolor,
    filecolor=linkcolor,
    urlcolor=linkcolor,
}
\usepackage{color,colortbl}
\definecolor{linkcolor}{rgb}{0.0,0.3,0.5}
\usepackage{tensind}
\tensordelimiter{?}
\DeclareGraphicsExtensions{.bmp,.png,.jpg,.pdf}
\usepackage{verbatim}
\usepackage[normalem]{ulem}
\usepackage{orcidlink}
\usepackage{soul}
\usepackage{multirow}

\usepackage[style=list]{glossaries}

\graphicspath{ {./figs/} }

\usepackage{newtxtext,newtxmath}

\usepackage[T1]{fontenc}

\usepackage{graphicx}	
\usepackage{amsmath, amssymb, graphicx, color}

\usepackage{xspace}

\makeglossaries
\glsdisablehyper
\newacronym{VMS}{VMS}{very massive star}
\newacronym{HD}{HD}{Humphreys-Davidson}
\newacronym{HR}{HR}{Hertzsprung-Russell}

\newacronym{MS}{MS}{main sequence}
\newacronym{ZAMS}{ZAMS}{zero-age main sequence}
\newacronym{TAMS}{TAMS}{terminal-age main sequence}
\newacronym{HG}{HG}{Hertzprung-gap}
\newacronym{CHeB}{CHeB}{core-helium burning}
\newacronym{AGB}{AGB}{asymptotic giant branch}
\newacronym{RGB}{RGB}{red giant branch}
\newacronym{BSG}{BSG}{blue supergiant}
\newacronym{YSG}{YSG}{yellow supergiant}
\newacronym{RSG}{RSG}{red supergiant}
\newacronym{WR}{WR}{Wolf-Rayet}
\newacronym{LBV}{LBV}{luminous blue variable}
\newacronym{BH}{BH}{black hole}

\newacronym{RLOF}{RLOF}{Roche-lobe overflow}

\newacronym{PPSN}{PPSN}{pair-instability pulsation supernova}
\newacronym{PSN}{PSN}{pair-instability supernova}

\newacronym{LVK}{LVK}{LIGO/Virgo/KAGRA}
\newacronym{GW}{GW}{gravitational wave}

\newacronym{MW}{MW}{Milky Way}
\newacronym{LMC}{LMC}{Large Magellanic Cloud}
\newacronym{SMC}{SMC}{Small Magellanic Cloud}

\newacronym{Mdot}{$\dot M$}{mass-loss rates}

\newacronym{LLM}{LLM}{Large Language Model}

\definecolor{burgundy}{rgb}{0.5, 0.0, 0.13}
\definecolor{coral}{rgb}{1.0, 0.5, 0.31}

\newcommand{\ST}[1]{{\color{coral}\textbf{[Paper II]}}}

\newcommand{\UCSD}{Department of Astronomy and Astrophysics, University of California, San Diego, La Jolla, CA 92093, USA}
\newcommand{\ITA}{Universit\"at Heidelberg, Zentrum f\"ur Astronomie (ZAH), Institut f\"ur Theoretische Astrophysik, Albert Ueberle Str. 2, 69120, Heidelberg, Germany}
\newcommand{\Padua}{Dipartimento di Fisica e Astronomia Galileo Galilei, Università di Padova, Vicolo dell’Osservatorio 3, I–35122 Padova, Italy}
\newcommand{\CAMK}{Nicolaus Copernicus Astronomical Center, Polish Academy of Sciences, ul. Bartycka 18, 00-716 Warsaw, Poland}
\newcommand{\PragueAstro}{Astronomický ústav, Akademie věd Ceské republiky, Fričova 298, 251 65 Ondřejov, Czech Republic}
\newcommand{\IAASARS}{IAASARS, National Observatory of Athens, 15236 Penteli, Greece}
\newcommand{\UAthens}{National and Kapodistrian University of Athens, 15784 Athens, Greece}

\defcitealias{Romagnolo_2026_AtlasII}{Atlas II}
\newcommand{\AtlasIIFirst}{\citet{Romagnolo_2026_AtlasII} (hereafter \citetalias{Romagnolo_2026_AtlasII})}
\newcommand{\AtlasII}{\citetalias{Romagnolo_2026_AtlasII}\xspace}
\newcommand{\AtlasIIp}{\citepalias{Romagnolo_2026_AtlasII}\xspace}

\begin{document}

\title{The stellar winds atlas I: Current uncertainties in mass-loss rates}

\author{Amedeo Romagnolo\orcidlink{0000-0001-9583-4339}}
\email{amedeoromagnolo@gmail.com}
\affiliation{\ITA}
\affiliation{\Padua}
\affiliation{\UCSD}
\affiliation{\CAMK}

\author{Floor S. Broekgaarden\orcidlink{0000-0002-4421-4962}}
\email{fbroekgaarden@ucsd.edu}
\affiliation{\UCSD}

\author{Konstantinos Antoniadis\orcidlink{0000-0002-3454-7958}}
\affiliation{\IAASARS}
\affiliation{\UAthens}

\author{Alex C. Gormaz-Matamala\orcidlink{0000-0002-2588-2391}}
\affiliation{\PragueAstro}


\begin{abstract}
Stellar winds are a major source of uncertainty in understanding the life and deaths of massive stars. 
Across studies in the field, prescriptions for stellar winds differ substantially in both their physical assumptions and implementation, making them a dominant contributor to model-to-model variation.
In this work, we present a systematic analysis of the physical assumptions underlying commonly adopted wind prescriptions for optically thin and optically thick winds of hot stars, as well as the winds of cool supergiants.
Our analysis reveals substantial discrepancies across all regimes: predicted mass-loss rates for optically thin winds differ by more than an order of magnitude, while rates for cool supergiants vary by several orders of magnitude, with even wider uncertainties arising in extrapolation regimes beyond the Humphreys-Davidson limit. These disparities introduce significant ambiguity into the predicted formation of Wolf-Rayet (WR) stars, a problem further compounded by the inconsistent application of transition criteria.
A central issue is the "cool Wolf-Rayet problem", a temperature regime where the classical electron-scattering Eddington factor ($\Gamma_{\rm e}$) loses physical consistency. Because this factor is widely used to determine WR mass-loss rates, its failure forces models to rely on uncertain extrapolations and ad-hoc corrections. 
We conclude that the dominant stellar wind uncertainties arise from a mismatch between the physical assumptions in stellar wind models and the structure of the stars to which they are applied. 
Our framework clarifies the origins of current theoretical discrepancies and identifies the key physical bottlenecks that must be addressed to improve mass-loss modeling for massive stars.
\end{abstract}


\maketitle

\section{Introduction}
\label{sec:intro}

The entire evolution of massive stars with \gls{ZAMS} masses $M_{\rm ZAMS}\gtrsim$\,8~$M_\odot$ is fundamentally shaped by mass loss through powerful stellar winds \citep{Meynet_1994,Eggenberger_2021,Garcia_2025}. 
These winds regulate the internal structure of massive stars---including their radial expansion \citep[e.g.,][]{Belczynski_2022_MK34,Romagnolo_2023,Gilkis_2025}---determine their final mass at core-collapse or pair-instability, and impact their galactic environments through mechanical feedback and chemical enrichment \citep{Maeder_1983,Dray_2003,Dray_2003b,Limongi_2006,Sander_2020,Farmer_2021,Martinet_2022,Higgins_2023,Josiek_2024}. 
Consequently, stellar evolution models rely on wind prescriptions or ``recipes'' to determine \gls{Mdot}, yet  the formulation and calibration of these recipes remain major challenges in stellar astrophysics.

For hot, massive stars, stellar winds are line-driven \citep{Lucy_70}, with mass-loss rates that depend  sensitively on luminosity and metallicity \citep{Abbott_1982,Pauldrach_1986,Kudritzki_1987}. These winds are broadly categorized as either \textit{optically thin} or\textit{ optically thick}. 
OB-type stars exhibit optically thin winds, in which momentum is transferred to the stellar material primarily through single-line scattering of photons from the photosphere with ions. 
As these stars evolve, their winds can transition to an optically thick regime, producing denser outflows and substantially higher mass-loss rates (from roughly 10$^{-9}$--10$^{-4.5}$ to 10$^{-5}$--10$^{-3}$~$M_\odot$ yr$^{-1}$; \citealt{Kudritzki_2000,GormazMatamala_2024a,Romagnolo_2024}) characteristic of \gls{WR} stars. The transition to thick winds (i.e. \gls{WR}) is still a subject of debate. Canonically, most 1D evolutionary models defined it as the evolutionary point at which the stellar surface would deplete its hydrogen abundance ($X_{\rm surf}$) below a specific threshold \citep[e.g.,][]{Glebbeek_2009,Ekstrom_2012}. More recently, such transition to thick winds has been updated to be a function of either the electron-scattering Eddington factor $\Gamma_{\rm e}$ \citep{Bestenlehner_2020}, the wind efficiency $\eta$ \citep{Sabhahit_2022}, which is directly correlated with the optical depth of the stellar photosphere (see for more details Sections~\ref{subsubsec:free-electron}, \ref{subsubsec:multi_scatter}, and \ref{subsubsec:new_eta}), or direct approximations of the winds' optical depth \citep[][see more in Section~\ref{subsubsec:tau_transition}]{AguileraDena_2022,Pauli_2022,Sen_2023}.

As massive stars between roughly 8 and 30~$M_\odot$ \citep{Meynet_2003} cool and expand into the \gls{YSG} and \gls{RSG} phases, the driving mechanism of their winds shifts away from line-driven, but the dominant physical process dominating mass loss in these cool supergiants remains poorly understood.
A cool supergiant wind interpretation involves dust-driven mass loss \citep{Reimers_1975}, enabled by dust formation at effective temperatures $T_{\rm eff}  \lesssim \,1.5$~kK \citep{Field_1974,Hofner_2018}. 
If dust can condense in the extended stellar atmosphere, its sufficiently high opacity would allow radiation pressure to efficiently expel material \citep{Hoyle_1962}. However, a major challenge is lifting gas high enough  for dust condensation to occur. 
Proposed mechanisms, such as strong near-surface convection and radial pulsations \citep{Yoon_2010}---similar to those in \gls{AGB} stars \citep{Hofner_2003,Neilson_2008}---have been shown to be insufficient for cool supergiants, which are hotter and exhibit smaller pulsation amplitudes than their \gls{AGB} counterparts \citep{ArroyoTorres_2015}. However, recent developments in pulsation-driven mass-loss scenarios propose that these mechanisms may indeed play a crucial role in aiding dust formation in the extended stellar atmosphere \citep[e.g.][]{Sengupta_2026}.

Alternative ideas have emerged in the field. 
Atmospheric turbulent pressure may be the primary driver of the outflow, with dust forming as a consequence rather than a cause \citep{Kee_2021}, reframing dust as a byproduct of the wind, rather than its primary cause, though it may subsequently enhance the outflow due to its high opacity. 
More recently, \citet{Fuller_2024} proposed a ``boil-off'' mechanism in which shocks from vigorous near-surface convection inject momentum into the atmosphere, producing mass-loss rates highly sensitive to the ratio of convective to escape velocities.
These competing physical models produce large uncertainties in existing mass-loss prescriptions, especially for \glspl{YSG}, where observational constraints are sparse, and for the most luminous supergiants approaching the \gls{HD} limit \citep{Humphreys_1979,Humphreys_1994}, where prescriptions require extrapolation far beyond empirically calibrated regimes.

Given these theoretical foundations, constructing universal,  physically consistent mass-loss prescriptions has proved challenging. 
Discrepancies between prescriptions grow particularly large for stars near the Eddington limit, in late evolutionary stages, or during phenomena not captured self-consistently in 1D stellar models—such as \gls{LBV} outbursts past the \gls{HD} limit.
%
The choice of a specific wind prescription is therefore far more than a technical input: it propagates through stellar models and can yield dramatically different evolutionary outcomes. 
This is apparent in the predictions of surface abundances \citep[e.g.][]{Martinet_2022}, stellar rotation rates \citep[e.g.][]{Limongi_2018}, and final fates \citep[e.g.][]{Renzo_2024}. 
At solar metallicity ($Z_\odot$)—an environment with rich observational constraints \citep{Martins_2005,Crowther_2010,Clark_2012}---recent studies have reported a wide range of final \gls{BH} masses owing to differences in adopted mixing and wind schemes \citep[e.g.][]{Bavera_2023,Gilkis_2024,Romagnolo_2024,Costa_2025,Hirschi_2025,Ugolini_2025}.
This divergence is driving renewed debate about the reliability of existing wind prescriptions and highlights  the urgent need for systematic comparison.


This work is the first part of the Stellar Winds Atlas series, where we present a pedagogical overview of several widely-used stellar wind prescriptions for isolated massive stars, placing our analysis within ongoing debates to understand and quantify the dominant sources of uncertainty in wind modeling\footnote{Our Atlas maps the diversity of mass-loss prescriptions in the literature, where we interpret this diversity as a practical uncertainty budget. For stellar atmospheres, this might not represent the full physical uncertainty ranges that would result from applying additional, model-specific constraints.}. By providing an unified framework for interpreting theoretical discrepancies in mass-loss rates, this work lays the foundations to explore how different wind recipes alter the evolution of massive stars. We directly apply this framework in \AtlasIIFirst, where we evaluate how these wind uncertainties impact the formation of black holes and their final masses at solar metallicity.
Our methods are described in Section~\ref{sec:method-atlas-paper-I}.
Section~\ref{sec:line-driven} examines line-driven winds and the assumptions behind recipes for both optically thin and optically thick regimes, with particular attention to the uncertain transition between them.
Section~\ref{sec:dust_driven} focuses on cool supergiant winds, comparing canonical and recent prescriptions and discussing challenges associated with the \gls{YSG} phase and extrapolations near the \gls{HD} limit. Finally, Section~\ref{sec:other_uncert} gives an overview of other sources of uncertainty that are not covered by the other sections.
Through this systematic comparison, we aim to provide a unified framework for interpreting theoretical discrepancies and identifying key bottlenecks in stellar-wind modeling.

\section{Method}
\label{sec:method-atlas-paper-I}

Our approach is a direct, side-by-side comparison of commonly-used stellar wind recipes, evaluated independently of the complexities inherent to full stellar evolution simulations. A list of stellar-wind prescriptions is provided in Table~\ref{tab:wind_acronyms}.
We compute mass-loss rates directly from the published analytical formulae, fitting functions, or tabulated data associated with each wind recipe. 

The prescriptions analyzed here, listed with their acronyms and sources in Table~\ref{tab:wind_acronyms}, are selected to represent a broad and representative sample of those wind models currently employed in massive star evolution and population synthesis studies. Our selection includes both foundational, widely-adopted recipes \citep[e.g.,][]{deJager_1988,NugisLamers_2000,Vink_2001} and more recent recipes that incorporate updated physics or new observational constraints \citep[e.g.,][]{Bestenlehner_2020,GormazMatamala_2023,Antoniadis_2024,Pauli_2025}. 
This curated set provides a comprehensive overview of the  landscape of theoretical uncertainties relevant to  massive-star modeling. We discuss the optically thin and thick winds models and present their comparisons in Section~\ref{sec:line-driven}, and similarly discuss the cool supergiant wind models and present their comparisons in Section~\ref{sec:dust_driven}.  A summary of recent stellar evolutionary models and their associated wind schemes is provided in Table~\ref{tab:models}.

\renewcommand{\arraystretch}{1.1}
\begin{table}[ht]
\centering
\caption{Overview of the stellar-wind prescriptions compared in this work.}
\label{tab:wind_acronyms}
\small 
\setlength{\tabcolsep}{6pt} 
\begin{tabular}{l l}
\hline
\textbf{ID} & \textbf{Reference} \\
\hline
\hline
\multicolumn{2}{c}{\textit{\gls{RSG} and red giants}}\\
\hline
Re75  & \citet{Reimers_1975} \\
\hline
\multicolumn{2}{c}{\textit{Line-driven}}\\
\hline
L89 & \citet{Langer_1989} \\
Ha95 & \citet{Hamann_1995}\textsuperscript{\textbf{a}} \\
Vb98 & \citet{vanbeveren_1998}\textsuperscript{\textbf{b}} \\
NL00  & \citet{NugisLamers_2000} \\
C01   & \citet{Crowther_2001} \\
V01   & \citet{Vink_2001} \\
EV06  & \citet{Eldridge_2006} \\
Y06   & \citet{Yoon_2006} \\
GH08  & \citet{Grafener_2008} \\
V11   & \citet{Vink_2011} \\
Z13   & \citet{Zdziarski_2013} \\
TSK16 & \citet{Tramper_2016}\\
V17   & \citet{Vink_2017} \\
Y17   & \citet{Yoon_2017} \\
S19   & \citet{Sander_2019}\\
Sh19  & \citet{Shenar_2019}\textsuperscript{\textbf{c}} \\
B20   & \citet{Bestenlehner_2020}\textsuperscript{\textbf{d}} \\
SV20  & \citet{Sander_2020} \\
VS21  & \citet{Vink_2021b} \\
Bj23  & \citet{Bjorklund_2023} \\
GM23  & \citet{GormazMatamala_2023} \\
K24   & \citet{Krticka_2024} \\
K25   & \citet{Krticka_2025}\textsuperscript{\textbf{e}} \\
P25   & \citet{Pauli_2025} \\
\hline
\multicolumn{2}{c}{\textit{Cool supergiants}}\\
\hline
dJ88  & \citet{deJager_1988}\textsuperscript{\textbf{f}} \\
NdJ90 & \citet{Nieuwenhuijzen_deJager_1990}\textsuperscript{\textbf{f}} \\
vL05  & \citet{vanLoon_2005} \\
Kee21 & \citet{Kee_2021}\\
Be23  & \citet{Beasor_2023} \\
Ya23  & \citet{Yang_2023} \\
A24   & \citet{Antoniadis_2024}\textsuperscript{\textbf{g}} \\
D24   & \citet{Decin_2024} \\
FT24 & \citet{Fuller_2024}\\
\hline
\multicolumn{2}{c}{\textit{\gls{LBV}}}\\
\hline
H00   & \citet{Hurley_2000} \\
Bk10  & \citet{Belczynski_2010} \\
LC18  & \citet{Limongi_2018}\\
\hline
\end{tabular}

\vspace{1ex}
\parbox{\columnwidth}{
\footnotesize \raggedright
\textsuperscript{a} With \citealt{Vink_2005} $Z$-dependence.\\
\textsuperscript{b} Different mass-loss formulations for thin, thick and cool supergiant winds.\\
\textsuperscript{c} Corrected tailored fits from \citealt{Shenar_2020}.\\
\textsuperscript{d} With \citealt{Brands_2022} $Z$-dependence.\\
\textsuperscript{e} {We adopted the more detailed fits from their \hyperlink{https://zenodo.org/records/15965163}{Zenodo data}}.\\
\textsuperscript{f} {Derived over a wide sample from O- to M-type stars, from supergiants to late-type red giants down to $\log (L/L_\odot)\approx3.3$. It can be technically used for any type of non-\gls{WR} and H-rich star, including line-driven winds.}\\
\textsuperscript{g} Corrected fits from \citealt{Antoniadis_2025_errata}.\\
}
\end{table}
\renewcommand{\arraystretch}{1.0} 

\begingroup
\renewcommand{\arraystretch}{1.4}
\begin{table*}
\centering
\caption{Atlas of unique massive-star evolution models from this work and the literature, summarizing the implemented wind recipes. \label{tab:models}}
\begin{tabular}{c|c|c c|c c|c}
\hline
\multicolumn{1}{c|}{\textbf{Model}} & \multicolumn{3}{@{}c@{}|}{\cellcolor{cyan!20}Line-driven (Section~\ref{sec:line-driven})} & \multicolumn{2}{@{}c@{}|}{\cellcolor{cyan!20}Cool supergiants (Section~\ref{sec:dust_driven})} & \multicolumn{1}{c}{\gls{LBV}} \\
\cline{2-6}
& \multicolumn{1}{c|}{Thin winds $\dot M$} & \multicolumn{1}{c}{Thick winds trans.} & \multicolumn{1}{c|}{$\dot M$} & \multicolumn{1}{c}{Transition $T_{\rm eff}$} & \multicolumn{1}{c|}{$\dot M$} & \\
\hline
Warm\_$\Gamma$ \AtlasIIp & P25 & $\eta$, $\Gamma_{\rm e}$\textsuperscript{a} & B20, SV20 & 10~kK & dJ88,A24 & -- \\
MSc \AtlasIIp & GM23, V01 & $\eta$>$\eta_{\rm trans}$\textsuperscript{a} & B20, SV20 & 10~kK & dJ88, A24 & -- \\
FESc$_{\rm A24}$ \AtlasIIp & GM23, V01 & $\Gamma_{\rm e}$\,>\,$\Gamma_{\rm e, trans}$\textsuperscript{a} & B20, SV20 & 10~kK & dJ88, A24 & -- \\
Dutch-ish \AtlasIIp & GM23, V01 & $X_{\rm surf}\,<\,$0.4 & NL00, EV06 & 10~kK & dJ88 & -- \\
Strong Winds Eta (SWE) &  V01 & $\eta$>$\eta_{\rm trans}$\textsuperscript{b} & V11 & 4~kK & dJ88 & -- \\
Strong Winds Gamma (SWG) &  V01 & $\Gamma_{\rm e}$\,>\,$\Gamma_{\rm e, trans}$\textsuperscript{a} & B20 & 10~kK & dJ88 & -- \\
KABS (Krticka, Antoniadis, Bestenlehner, Sander) & K24  & $\eta$, $\Gamma_{\rm e}$\textsuperscript{a} & B20, SV20 & 10~kK & dJ88, A24 & -- \\
Hydrodynamic cool winds (HyD-Cool) &  K24 & $\Gamma_{\rm e}$\,>\,$\Gamma_{\rm e, trans}$\textsuperscript{a} & B20, SV20 & 4.5~kK & FT24 & -- \\
\hline
\cite{GormazMatamala_2026} & GM23, V01 & $X_{\rm surf}$, $\Gamma_{\rm e}$\textsuperscript{a} & B20, SV20 & 10~kK & dJ88, Ya23 & -- \\
\cite{Pauli_2026} & P25 & $\Gamma_{\rm e}$\,>\,$\Gamma_{\rm e, trans}$ & Sh19, P25 & 10~kK\textsuperscript{c} & Ya23 & Original \\
\cite{Costa_2025} &  V01, GH08, V11 & $X_{\rm surf}\,<\,$0.3 & S19\textsuperscript{d} & 10~kK & dJ88\textsuperscript{e} & -- \\
\cite{Gilkis_2025} &  VS21, V17 & $\Gamma_{\rm e}$\,>\,$\Gamma_{\rm e, trans}$ & GH08, Sh19, SV20 & 22~kK & dJ88, VS21 & -- \\
\cite{GormazMatamala_2025} & GM23, V01 & $X_{\rm surf}$, $\Gamma_{\rm e}$\textsuperscript{a} & B20, SV20 & 10~kK & dJ88 & -- \\
\cite{Hirschi_2025} &  V01\textsuperscript{f} & $X_{\rm surf}\,<\,$0.3 & NL00, EV06, GH08\textsuperscript{f} & 10$^{\rm 3.9}$~K & dJ88\textsuperscript{f} & -- \\
\cite{Jin_2025} & V01 & $X_{\rm surf}\,<\,$0.7 & NL00, Y17 & 25~kK & NdJ90, V01 & -- \\
\cite{Keszthelyi_2025} &  B20 & $X_{\rm surf}\,<\,$0.4 & SV20 & 10~kK & dJ88, vL05 & -- \\
\cite{Romagnolo_2025} & 0.5$\times$V01 & $X_{\rm surf}\,<\,$0.4 & 0.5$\times$NL00 & 10~kK & 0.5$\times$dJ88 & -- \\
\cite{Ugolini_2025} &  V01 & $X_{\rm surf}\,<\,$0.4 & NL00 & 12~kK & dJ88, vL05 & LC18 \\
\cite{Cheng_2024} & 0.8$\times$V01 & $X_{\rm surf}\,<\,$0.4 & 0.8$\times$NL00 & 4~kK & D24  & Original\\
\cite{Josiek_2024} & V01\textsuperscript{f} or B20\textsuperscript{f} & $\Gamma_{\rm e}$\,>\,$\Gamma_{\rm e, trans}$\textsuperscript{a} & NL00, EV06, GH08\textsuperscript{f} & 10~kK & dJ88, C01\textsuperscript{f} & -- \\
\cite{Romagnolo_2024} & GM23, V01 & $X_{\rm surf}$, $\Gamma_{\rm e}$\textsuperscript{a} & B20, SV20 & 10~kK & dJ88 & -- \\
\cite{Sabhahit_2023} & V01 & $\eta$>$\eta_{\rm trans}$\textsuperscript{b, g} & V11, SV20 & 4~kK & dJ88 & --\\
\cite{Keszthelyi_2022} & 0.5$\times$V01 & $X_{\rm surf}\,<\,$0.4 & NL00 & 10~kK & vL05 & -- \\
\cite{Klencki_2022} & V01, NdJ90 & $\Gamma_{\rm e}$\,>\,$\Gamma_{\rm e, trans}$ & Y06 & 10~kK & dJ88 & --\\
\cite{Eggenberger_2021} &  V01\textsuperscript{f} & $X_{\rm surf}\,<\,$0.3 & NL00, EV06, GH08\textsuperscript{f} & 10$^{\rm 3.9}$~K & dJ88\textsuperscript{f} & -- \\
\cite{Agrawal_2020} & V01\textsuperscript{h} & $X_{\rm surf}\,<\,$0.4 & NL00\textsuperscript{h} & 10~kK & dJ88\textsuperscript{h} & --\\
\hline
\rule{0pt}{2.5ex}Dutch winds & V01 & $X_{\rm surf}\,<\,$0.4 & NL00 & 10~kK & dJ88 & -- \\
\hline

\multicolumn{7}{@{}c@{}}{\cellcolor{lightgray}\rule{0pt}{2.5ex}Population synthesis codes\rule[-1.2ex]{0pt}{0pt}}\\

\hline
\rule{0pt}{2.5ex}{\tt StarTrack} \citep{Belczynski_2010} &  V01 or NdJ90 & H00 & Ha95 & 12.5~kK & NdJ90 & Original\rule[-1.2ex]{0pt}{0pt}\\
{\tt COMPAS} \citep{Merritt_2025} & VS21 &  $\eta$>$\eta_{\rm trans}$\textsuperscript{i} & V11, SV20 & 8~kK & D24 & Bk10 \\
{\tt SeBa} \citep{Dorozsmai_2024} & V01, NdJ90 & H00 & SV20 & 8~kK & Re75, NdJ90 & Bk10\\
{\tt BPASS} \citep{Eldridge_2017} & V01 & $X_{\rm surf}\,<\,$0.4 & NL00 & 10~kK & dJ88 & -- \\ 
{\tt BINARY\_C} \citep{Hendriks_2023} & V01 & H00 & Y07 & 11~kK & Re75 & H00\\
{\tt COSMIC} \citep{Breivik_2020} & V01 & H00 & Ha95 & 12.5~kK & NdJ90 & Bk10\\
{\tt COSMIC-METISSE} \citep{Maclean_2026} & K25 & $X_{\rm surf}\,<\,$0.4 & NL00 & 10~kK & D24 & Bk10\\
{\tt MOBSE} \citep{Giacobbo_2018} & V01 & H00 & Ha95 & 12.5~kK & NdJ90\textsuperscript{e} & Bk10\textsuperscript{d}\\
{\tt NBODY7/BSE} \citep{Banerjee_2020} & V01 & H00 & Ha95 & 12.5~kK & NdJ90 & Bk10\\
{\tt MOCCA} \citep{Kamlah_2022} & V01 & H00 & Ha95 & 12.5~kK & NdJ90 & Bk10\\
{\tt TRES}\textsuperscript{j} \citep{Toonen_2016} & V01, NdJ90 & H00 & SV20 & 8~kK & Re75, NdJ90 & Bk10\\
{\tt POSYDON}\textsuperscript{j} v2 \citep{Andrews_2024} &  V01 & $X_{\rm surf}\,<\,$0.4 & NL00 & 10~kK & dJ88 & Bk10\\
{\tt SEVN}\textsuperscript{j} \citep{Iorio_2023} &  V01, GH08, V11 & $X_{\rm surf}\,<\,$0.3 & S19\textsuperscript{d} & 12~kK & dJ88\textsuperscript{e} & -- \\
\hline
\end{tabular}
\parbox{\textwidth}{\footnotesize
\textsuperscript{a}{For cool \gls{WR} stars ($T_{\rm eff}\,<\,$30~kK), the optically thin mass loss is still applied.}\\
\textsuperscript{b}{For decreasing $\Gamma_{\rm e}$ or $\Gamma_{\rm e}$ values lower than the one at thick winds transition, the optically thin mass loss is still applied.}\\
\textsuperscript{c}{The transition to cool supergiant winds also requires that the cool layers to be convective account for more than 0.1\% of the total hydrogen envelope mass.}\\
\textsuperscript{d}{With $Z$-dependence of \citealt{Costa_2021}, derived with fits from \citealt{Vink_2015}.}\\
\textsuperscript{e}{Metallicity-dependent mass loss calibration based on \cite{Vink_2001}}\\
\textsuperscript{f}{$3\times\dot M$ for supra-Eddington surfaces \citep{Ekstrom_2012}.}\\
\textsuperscript{g}{Calculation of $\eta$, $\eta_{\rm trans}$, and $\Gamma_{\rm e}$ under assuming that the star is chemically homogeneous (see Section~\ref{subsubsec:multi_scatter}).}\\
\textsuperscript{h}{Applies 10\% of the \cite{Paczynski_1986} multiplier to enhance mass loss if stellar luminosity is greater or equal to 110\% of the Eddington luminosity.}\\
\textsuperscript{i}{Calibrations for population synthesis from \cite{Merritt_2025}.}\\
\textsuperscript{j}{These codes can interpolate detailed evolutionary simulations. Here are shown only the default stellar tracks/models presented with each code.}\\
\textbf{Top panel}: new models from this study (see Section~\ref{subsubsec:wind_implem}) and selected ones from \AtlasII.
\\\textbf{Middle panel}: models from a selected recent literature sample. The Dutch winds are shown separately as they represent the most widely used models in the literature.
\\\textbf{Bottom panel}: population synthesis models. 
}
\end{table*}\
\renewcommand{\arraystretch}{1.0} 
\endgroup
We evaluate the mass-loss rates with the analytical formulations from each presented wind models within a set parameter space at solar metallicity, $Z_\odot$\,=\,0.0142 \citep{Asplund_2009}, a regime where  mass loss is strong and Milky Way massive stars offer observational constraints \citep[e.g.,][]{Martins_2005,Crowther_2010,Clark_2012}. 
This parameter choice provides a  well-studied, uniform baseline that allows a clean  comparison across recipes. 
When a prescription's original calibration does not span the full parameter space required for our comparison, we apply extrapolations and suggest alternative calibration options.
\subsection{{\tt MESA} models}
To demonstrate how certain physical criteria depend on the underlying stellar structure, we compute a dedicated grid of stellar models  using Modules for Experiments in Stellar Astrophysics~\citep[\texttt{MESA}][]{Paxton_2011,Paxton_2013,Paxton_2015,Paxton_2018,Paxton_2019,Jermyn_2023}, version 24.08.1. 

\paragraph{Input Physics}
We adopt the Ledoux criterion \citep{Ledoux_1947} for convective boundaries with a mixing length parameter of $\alpha_{\rm MLT} = 1.82$ and exponential overshooting with a parameter $f_{\rm ov} = 0.05$. 
We include rotation with an initial ratio of angular velocity to critical velocity of $\Omega/\Omega_{\rm crit} = 0.4$. Rotational mixing and associated mass-loss enhancement \citep{Friend_1986,Langer_1998} follow the calibration of \citet{Heger_2000}, while magnetic angular momentum transport is modeled via the Tayler-Spruit dynamo \citep{Tayler_1973,Spruit_2002,Heger_2005}.

\subsubsection{Wind Scheme Implementations}
\label{subsubsec:wind_implem}

To evaluate the model dependence of transition criteria (see Section~\ref{subsec:thick_winds}) and explore variability across different physical assumptions (see Section~\ref{subsec:calibrations}), we implemented the following wind configurations:

\begin{itemize}
    \item Standard Dutch wind scheme: We use this case to explore variability with metallicity calibrations and rotation, serving as a standard baseline due to its wide use in research papers \citep{Glebbeek_2009}.
    \item Strong Winds Eta (SWE): We use this case to represent a strong mass-loss scenario. This configuration applies \citet{Vink_2001} for thin and low-$\Gamma_{\rm e}$ thick winds (see more in Section~\ref{subsubsec:multi_scatter}). It applies \citet{deJager_1988} for $T_{\rm eff}<4$~kK, \citet{Vink_2011} for high-$\Gamma_{\rm e}$ thick winds, and \cite{Sander_2020} for $T_{\rm eff}>10^2$~kK. This setup modifies the original \cite{Sabhahit_2023} framework by incorporating the dynamic efficiency calculations described in Section~\ref{subsubsec:new_eta}.
    \item Strong Winds Gamma (SWG) model: We use this case to represent a strong mass-loss scenario based on the $\Gamma_{\rm e}$ transition to thick winds (Section~\ref{subsubsec:free-electron}). \cite{Vink_2001} for thin winds and cool \glspl{WR}, \cite{Bestenlehner_2020} ($X_{\rm surf}<10^{-7}$) or \citet{Sander_2020} ($X_{\rm surf}\geq10^{-7}$) for warm \glspl{WR}, \cite{deJager_1988} at $T_{\rm eff}<10$~kK, and $\Gamma_{\rm e} \geq 0.5$ transition to thick winds.
    \item KABS (Krticka, Antoniadis, Bestenlehner, Sander) winds: \cite{Krticka_2024} for thin winds\footnote{At $Z\lesssim0.2Z_\odot$ we recommend the use of \cite{Krticka_2025}, since the driving elements for winds are different and mass-loss rates follow a different behavior.} and \glspl{WR} at $T_{\rm eff}<30$~kK, \cite{Bestenlehner_2020} ($X_{\rm surf}\geq10^{-7}$) or \citet{Sander_2020} ($X_{\rm surf}<10^{-7}$) for warm \glspl{WR}, \cite{Antoniadis_2024} at $T_{\rm eff}\leq4$~kK, \cite{deJager_1988} at $T_{\rm eff}\leq10$~kK, and transition to thick winds as $\eta \geq \eta_{\rm trans}$ or $\Gamma_{\rm e} \geq 0.5$ (Section~\ref{subsubsec:free-electron}), whatever comes first.
    \item HydroDynamic Cool winds (HyD-Cool): We use this model to show the mass-loss rates from the hydrodynamic-approximation model from \citet{Fuller_2024}, since it does not possess a direct and univocal dependence on the global stellar surface quantities \textit{L}, $T_\text{eff}$, \textit{Z}. In this model, we use \cite{Krticka_2024} for thin winds and \glspl{WR} at $T_{\rm eff}<30$~kK, the $\Gamma_{\rm e} \geq 0.5$ transition to optically thick winds, \cite{Bestenlehner_2020} ($X_{\rm surf}\geq10^{-7}$) or \citet{Sander_2020} ($X_{\rm surf}<10^{-7}$) for warm \glspl{WR}, and \cite{Fuller_2024} at $T_{\rm eff}<4.2$~kK, which is the original designed threshold for \gls{RSG} winds in the original paper. We do not extend the cool supergiant winds at higher temperatures because this model was only calibrated for \glspl{RSG}, since warmer objects, at least below the \gls{HD} limit, do not develop massive convective envelopes.

\end{itemize}

\subsubsection{Automated stellar winds tools}

\paragraph{\textbf{{\tt Argus} LLM-driven {\tt MESA} inlist generator}} an automated, \gls{LLM}-driven builder for the Stellar Winds Atlas. While standard reproduction packages provide static files, {\tt Argus} introduces a dynamic approach to modeling. It consists of a specialized context dictionary designed for \glspl{LLM}. This file encodes the structural logic of our {\tt MESA} \textit{inlist} and \textit{run\_star\_extras} files, which are able to initiate stellar evolution simulations with any of the winds prescriptions from Table~\ref{tab:wind_acronyms}, alongside their respective recommended parameter spaces.

By uploading this dictionary alongside our default \textit{inlist\_project} to an \gls{LLM}, users can interact with the Atlas via a natural language interface. {\tt Argus} allows users to generate valid, complex {\tt MESA} Atlas winds configurations by simply requesting specific physical scenarios (e.g., "Configure a model using the KABS wind scheme at LMC metallicity, but use the oldest cool supergiant winds formulation you have for red supergiants"). The dictionary contains the physical limitations of each recipe, {\tt Argus} can automatically warn users if their requested parameters fall outside calibrated regimes. 

{\paragraph{\textbf{Wind Atlas automated mass-loss comparator}}
In order to dynamically compare different mass-loss recipes as a function of metallicity, luminosity, and effective temperature, as well as to visualize their variability on \gls{HD} diagrams, we developed a web application for our Atlas paper series, available at \url{https://amedeorom.github.io/Stellar_Winds_Atlas/}.
}
    
\section{Line-Driven Winds}
\label{sec:line-driven}
%

\begin{figure}[!ht]
\centering
\includegraphics[width=1\columnwidth]{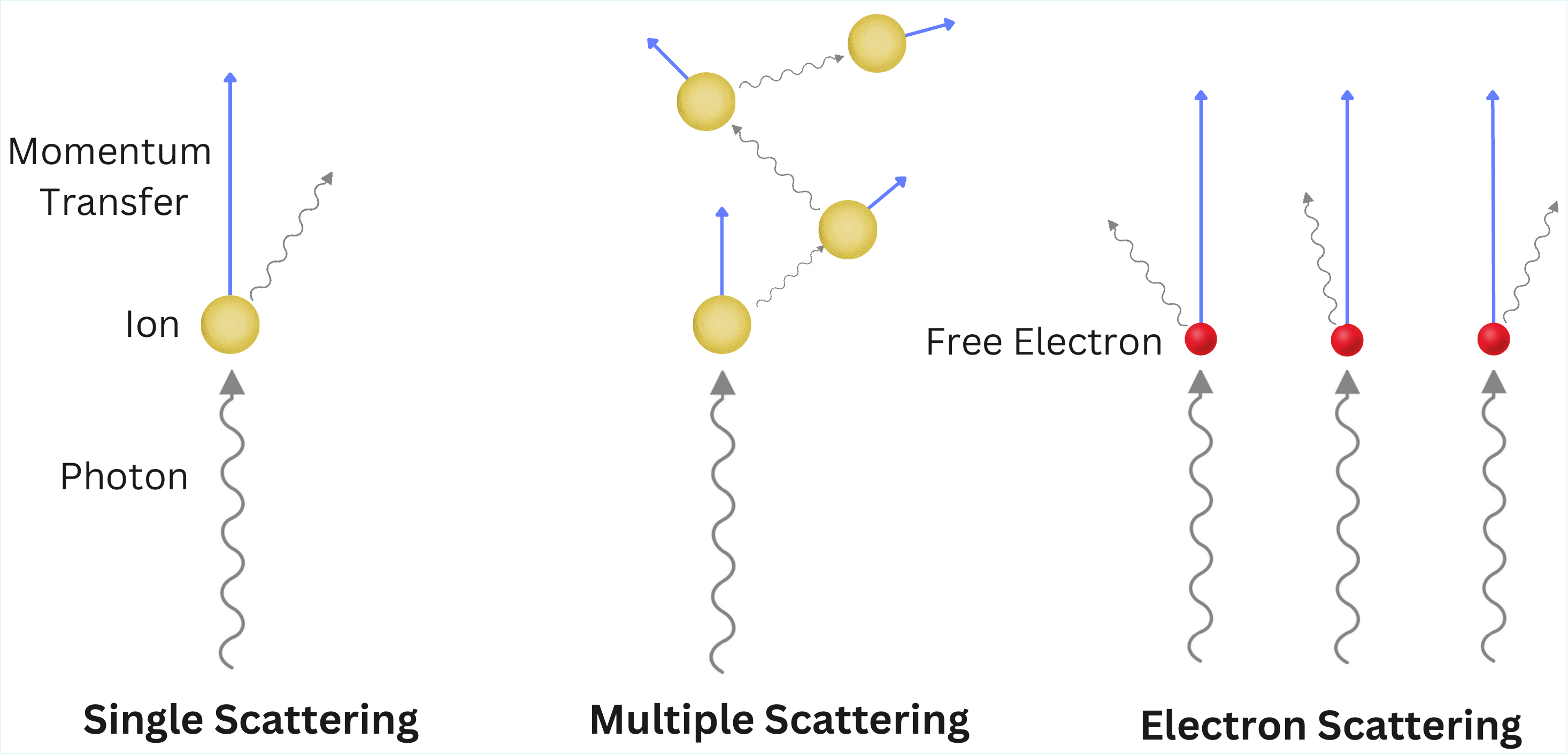}
\caption{Schematic representation of the primary physical mechanisms governing optically thin and optically thick winds. 
All mechanisms operate to some extent in stellar atmospheres, but their relative contributions differ. 
In optically thin winds, single scattering processes dominate, whereas in optically thick winds, multiple scattering (i.e., deposit of more angular momentum) and free-electron (Thomson) scattering processes become the main drivers.}
\label{fig:line_driven}
\end{figure}

Throughout most of their lives, wind of massive stars is dominated by radiatively-driven mechanism: the strong radiation (mostly ultraviolet) coming from the stellar photosphere transfers momentum to the particles composing the wind \citep{Puls_2008}.
This momentum transfer is done by both electron scattering, or by scattering over ions absorbing and re-emitting photons at certain line transitions (reason why we also talk about ``line-driven winds'').
See Fig.~\ref{fig:line_driven} for a full scheme.
Line-driven mechanism is strongly metallicity-dependent, as heavier elements provide a large number of spectral lines capable of absorbing photons.

At the beginning of their lives, the line-driven wind of OB-type stars is optically thin.
This means, the stellar radius is well-defined at an optical depth of $\tau$\,=\,2/3, with a subsonic velocity profile just above the surface. 
For more evolved stages, we found scenarios where the wind becomes so dense that a clear separation between the photosphere and the expanded atmosphere disappears, thus generating denser outflows with substantially higher mass-loss rates (than during the OB phase), characteristic of \gls{WR} stars \citep{Crowther_2007,Grafener_2008}.
Because of this higher density, stratification of the wind becomes more uncertain and therefore stellar radii and effective temperatures are poorly constrained---an issue known as the ``WR radius problem'' \citep{Langer_1988}. 
%

\subsection{Optically thin winds}
Figure~\ref{fig:thin} shows our overview of thin winds.
The most widely used prescription of mass loss in this regime comes from \citet{Vink_1999,Vink_2000,Vink_2001}, which is one of the main components of the widely-used Dutch winds model.
However, this recipe does not include wind clumping in OB stars, a phenomenon caused by radiative instabilities and currently widely observed in OB stars \citep{Bouret_2005,Lai_2022,Lai_2024,Verhamme_2024}. 
More recent approaches that explicitly account for wind clumping predict mass-loss rates reduced by factors 2--5 compared to older models \citep{Bouret_2005,Surlan_2012,Surlan_2013,Smith_2014,Sander_2017,GormazMatamala_2021,Pauli_2025}, which is also observationally justified by the fact that rotating stars were shown to not considerably spin down during \gls{MS}, suggesting therefore limited angular momentum ejection, hence weaker winds than canonically predicted \citep{Nathaniel_2025b}. 
Despite this progress, both theoretical and observational constraints on clumping \citep{Verhamme_2024,BerniniPeron_2025} and overall mass-loss rates \citep{Backs_2024} remain sufficiently wide that the classical \citet{Vink_1999,Vink_2000,Vink_2001} prescription cannot be fully ruled out across all parameter spaces.

An important caveat arises for stars at an effective temperature $T_{\rm eff} \lesssim 25$~kK. In its formulation, \cite{Vink_2001} incorporated the concept, originally proposed by \cite{Pauldrach_1990} and supported observationally by \cite{Lamers_1995} and \cite{Markova_2008}, of a bistability jump, where  Fe~III  recombination increases opacity and enhances wind-driven mass loss rates by an order of magnitude. 
However, recent studies \citep{Krticka_2021,Bjorklund_2023,BerniniPeron_2024,deBurgos_2024,Verhamme_2024,Alkousa_2025,Hawcroft_2026,Verhamme_2026} find little evidence for such a mass-loss enhancement, restricting the applicability of this recipe to a subset of hotter stars. 
Some recent work  \citep{Krticka_2025, BerniniPeron_2023, Krticka_2024} supports a weaker or metallicity-dependent bistability jump, though the onset temperature that they find varies.

\subsubsection{Mass-loss rates}
\begin{figure*}[!ht]
\centering
\includegraphics[width=0.99\textwidth]{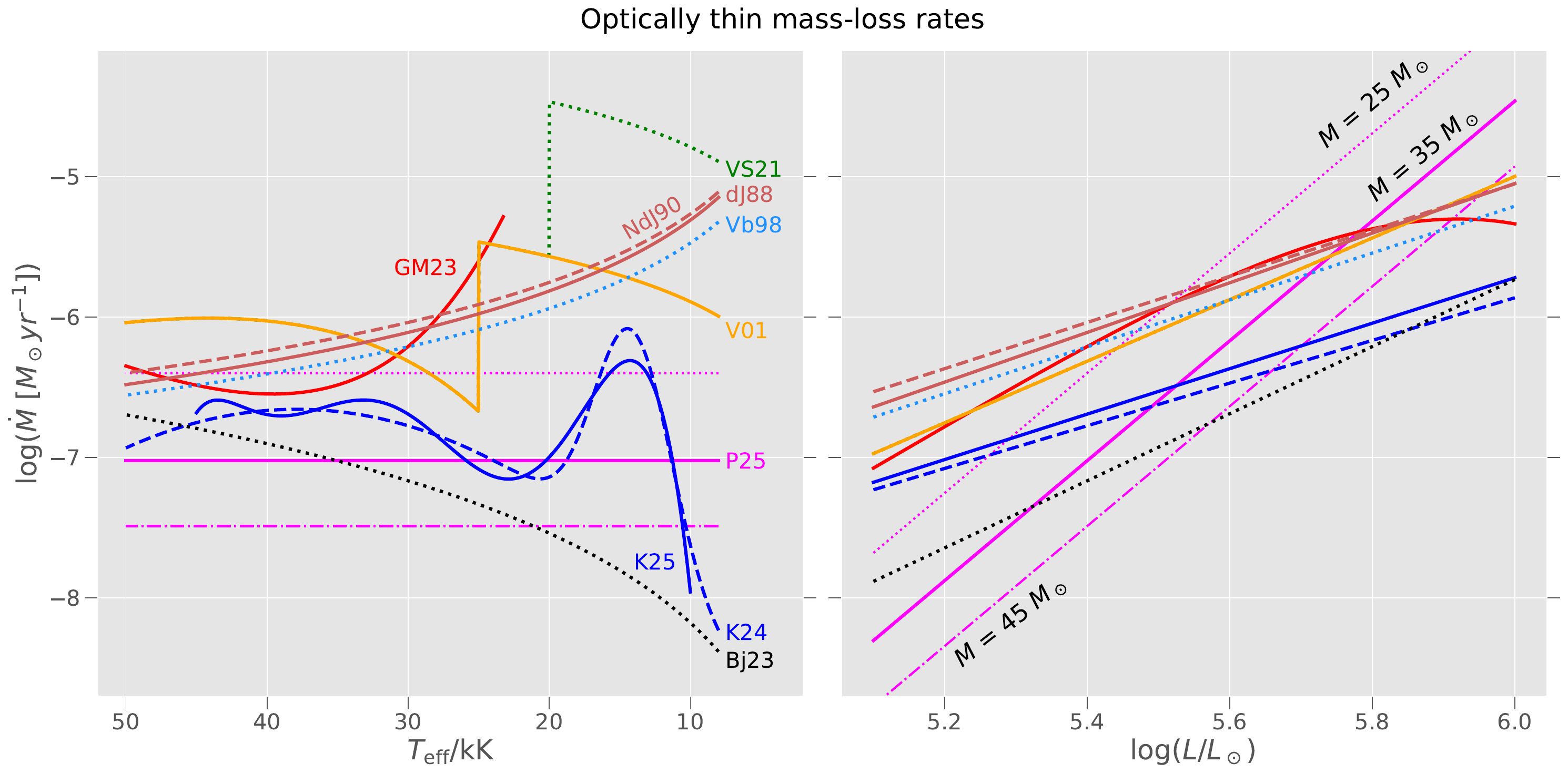}
\caption{Optically thin mass-loss rates for wind recipes from the literature (acronyms listed in Table~\ref{tab:models}). 
Left: mass-loss rates versus effective temperature for a 35~$M_\odot$ star and $\log(L/L_\odot)$\,=\,5.4 at $Z_\odot$ . 
Right: mass-loss rates versus luminosity for the same star at $T_{\rm eff}$\,=\,30~kK. 
The vertical dotted line marks the \gls{HD} limit.
\cite{Pauli_2025} is independent of $T_{\rm eff}$, and rates for this recipe are shown for masses of 25, 35, and 45~$M_\odot$. Rates from \cite{GormazMatamala_2023} are shown only for $\log g > 3$, the limit of their stated applicability for extrapolation. R26 is shown for relative solar abundances only.
\cite{Krticka_2025} was only applied within the $T_{\rm eff}\in[10; 45]$~kK range, as it leads to consideable underestimates in mass loss from extrapolation past its original parameter space.
Predictions for the mass-loss rates differ by orders of magnitude, especially for $T_{\rm eff}<25$~kK among models that include at least one bistability jump. 
}
\label{fig:thin}
\end{figure*}
In Fig.~\ref{fig:thin} we show our overview of the commonly-used mass-loss rates for a $Z_\odot$ 35~$M_\odot$ star with $\log(L/L_\odot)$\,=\,5.4 across 8~kK (the \gls{YSG} threshold) and 50~kK.
We find that the \cite{Vink_2001}, \cite{vanbeveren_1998}, \cite{Nieuwenhuijzen_deJager_1990}, and \cite{deJager_1988} recipes yield the strongest stellar winds\footnote{\cite{deJager_1988} and \cite{Nieuwenhuijzen_deJager_1990} were derived from the same observational sample, with \cite{vanbeveren_1998} updating 24 of those stars from \cite{Puls_1996}, hence the similar mass-loss rates.}, with the bistability jump producing an increase of more than an order of magnitude. Evolutionary modeling shows that such strong mass loss leads to an early-formed WNh population, whereas weaker, monotonic prescriptions such as \cite{Bestenlehner_2020} produce late-formed WR stars that pass through a cool-supergiant phase before stripping (\citealt{Josiek_2024}, \AtlasII).
Including the second bistability jump from \citet{Vink_2021b} can further boost mass-loss rates by another order of magnitude as shown in Figure~\ref{fig:thin}, though this feature remains unverified observationally \citep{deBurgos_2024,Verhamme_2024,Verhamme_2026}. 
Instead, recent observations suggest that the Eddington parameter $\Gamma_{\rm e}$, rather than $T_{\rm eff}$, is the primary driver of enhanced mass loss \citep{Pauli_2025}. 
Consequently,  population synthesis codes adopting  \cite{Nieuwenhuijzen_deJager_1990}, \cite{vanbeveren_1998}, \cite{Vink_2001}, or \cite{Vink_2021b} likely overestimate mass-loss rates in the optically thin winds regime,  accelerating the depletion of hydrogen (H)-rich envelopes and leading to premature formation of \gls{WR} stars.

\subsection{Optically thick (Wolf-Rayet) winds}
\label{subsec:thick_winds}
The mass-loss rates of optically thick winds remain highly uncertain, both theoretically and observationally, as large samples of observed \gls{WR} stars are still limited \citep{Shenar_2024}. We show an overview of thick wind recipes in Figure~\ref{fig:thick}.
Current observations indicate at least three distinct \gls{WR} populations with different mass-loss rates 
\citep[\gls{WR} classifications from][]{Yusof_2013,Aadland_2022,Martinet_2023}
\begin{enumerate}[label=\roman*)]
  \item Hydrogen (H)-rich WNL and WNh stars \citep{Vink_2011,Bestenlehner_2014,Grafener_2021}
  \item Surface H($X_{\rm surf}$)-poor and C/O-poor ($^{12}{\rm C}$+$^{16}{\rm O}\,\leq\,0.03$\,$^4$He) WNE stars
  \item H-free ($X_{\rm surf}\,<\,{\rm 10^{-7}}$) and C/O rich WC and WO stars, including transitional WN/WO phases \citep{Sander_2025}.
\end{enumerate}
Beyond the absolute mass-loss rates, a key uncertainty concerns the physical criterion that governs the onset of optically thick winds, which is connected with the physical origin of WR stars.
Stellar evolution models commonly adopt a surface hydrogen threshold, $X_{\rm surf}$, to trigger this transition \citep[e.g.][]{Glebbeek_2009,Ekstrom_2012,Eggenberger_2021,Romagnolo_2023,Cheng_2024,Costa_2025,Hirschi_2025,Romagnolo_2025,Ugolini_2025}, with values such as $X_{\rm surf}<0.4$ used to represent thick-wind phases in post-\gls{YSG}, post-\gls{RSG}, and post-\gls{LBV} stars \citep{Kohler_2015}.
This criterion is designed to perform the 'classical' \gls{WR} stars which are formed by the earlier removal of their outer envelope, but it fails to explain H-rich WNh stars \citep[e.g.][]{Bestenlehner_2014,Tehrani_2019,Martins_2022}, as well as both low-mass \citep{Wang_2017,Wang_2018,Bodensteiner_2020,Irrgang_2020,Wang_2021,ElBadry_2022b} and intermediate-mass \citep{Drout_2023,Gotberg_2023} H-poor stripped stars that do not show \gls{WR} characteristics. 

This mismatch reflects the historical calibration of  evolutionary models on classical, H-depleted late-type \gls{WR} stars---the first of such objects observed---rather than on the full diversity of massive stars.
In contrast, the most massive stars can display \gls{WR}-like spectra, with strong emission lines and  boosted mass-loss rates, from the onset of core hydrogen burning, when they are still H-rich. 
Thus, the physical boundary between optically thin and thick winds remains a topic of active debate. 
Recent hydrodynamic modeling of B hypergiants \citep{BerniniPeron_2025} illustrates this complexity: these stars can display strong emission lines (typically interpreted as signatures of dense winds) while still possessing optically thin outflows. 
This decoupling between spectral appearance and (true) wind optical depth underscores the difficulty of establishing a reliable physical criterion for the emergence of true optically thick, \gls{WR}-like winds.


\subsubsection{Free electron (Thomson) scattering transition}
\label{subsubsec:free-electron}
The strength of \gls{WR} winds is closely related to the Eddington factor, $\Gamma_{\rm e}$, which measures the ratio of radiative to gravitational forces. 
As a star approaches the Eddington limit ($\Gamma_{\rm e}\rightarrow1$), its mass-loss rate \gls{Mdot} increases steeply \citep{Vink_2011,Vink_2012}. 
%
Following the classical picture of \cite{Eddington_1926}, in which free-electron (Thomson) scattering  dominates the opacity, the electron Eddington factor is  
\begin{equation}\label{eq:Gamma_e}
    \Gamma_{\rm e}=\frac{L\kappa_{\rm e}}{4\pi cGM}
\end{equation}
with $\kappa_\text{e}$ the electron-scattering opacity. 
For a fully ionized plasma, this can be written  (in base-10 logarithm) as 
\begin{equation}\label{eq:logGammaEdd}
    \log\Gamma_\text{e}=-4.813+\log(1+X_\text{surf})+\log \frac{L}{L_\odot}-\log \frac{M}{M_\odot}
\end{equation}
%
Because of its simplicity,  $\Gamma_{\rm e}$ is frequently used in stellar evolution codes as a switch between thin and thick wind regimes. 
\citet{Bestenlehner_2020} extended the theoretical $\dot M-\Gamma_\text{e}$ relation from the CAK theory introduced in \cite{Castor_1975_cak} toward the Eddington limit, showing that the slope of the relation changes significantly between $\Gamma_\text{e}\ll1$ and $\Gamma_\text{e}\rightarrow1$.
Importantly, the transition point $\Gamma_\text{e,trans}$ is relatively robust to uncertainties in stellar atmosphere models: for stars with masses between 25 and 120~$M_\odot$ and metallicities between $Z$\,=\,0.006 (LMC) and $Z$\,=\,0.0142 (solar) \citep{GormazMatamala_2022b}---$\Gamma_{\rm e,trans}$ remains nearly constant at $\approx 0.48$. 
Consequently,  many stellar evolution models adopt $\Gamma_{\rm e}$\,=\,0.5 as an approximate thin-thick wind transition for this metallicity regime \citep{Romagnolo_2024}, adequately predicting the HR diagram position of WNh stars with $X_\text{surf}>0.3$ in the Milky Way \citep{GormazMatamala_2025}.

\paragraph{Model uncertainties} This transition condition cannot be  computed self-consistently in evolutionary models and must instead be calibrated using observations and stellar atmosphere simulations \citep[see e.g.][]{Bestenlehner_2020}. 
At higher metallicities, free-electron scattering is not the primary driver of optically thick winds; rather, $\Gamma_{\rm e}$ only acts as a proxy for Fe-opacity beneath the stellar surface. 
Because the iron (Fe) opacity peak shifts with metallicity, $\Gamma_{\rm e,trans}$ is not universal. 
\cite{Grafener_2008} showed that $\Gamma_{\rm e,trans}$ scales inversely with $Z$, although their calibration differs from more recent work \citep{Bestenlehner_2020}. 
Thus, while $\Gamma_{\rm e,trans}\approx0.5$ is appropriate for solar to LMC metallicities, it should increase towards the lower $Z$, reflecting the weaker Fe opacity rather than representing a fixed physical threshold. We further note that \cite{Sabhahit_2022} showed that this transition point is largely insensitive to some CAK model uncertainties, which limits its dynamic response to the specific details of the wind-driving model. This means that such a transition condition functions more as a global structural indicator of the L/M ratio than as a dynamic response to the underlying wind-driving physics. Consequently, the $\Gamma_{\rm e}$ criterion should be interpreted as a diagnostic proxy for identifying stars near the Eddington limit within 1D frameworks, rather than a resolved physical threshold derived from first principles.

\subsubsection{Multiple scattering transition}
\label{subsubsec:multi_scatter}

At high metallicities, multiple scattering becomes the primary driver of optically thick winds. A self-consistent way to link this process to the onset of \gls{WR} winds is based on the efficiency of its radiative momentum transfer \citep{Grafener_2017, Grafener_2021}. 
Following \cite{Grafener_2017}, this efficiency can be quantified with the optical depth at the sonic point $\tau_{\rm s}$ and the wind efficiency parameter $\eta$,
\begin{equation}\label{eq:opticaldepth}
    \tau_\text{s}\simeq\frac{\dot M \varv_\infty}{L/c}\left(1+\frac{\varv_\text{esc}^2}{\varv_\infty^2}\right)
\end{equation}
\begin{equation}\label{eq:eta}
    \eta\equiv\dot M\varv_\infty/(L/c)
\end{equation}
where $\varv_\infty$ is the terminal velocity. 
When $\eta$ exceeds the threshold 
\begin{equation}\label{eq:eta_trans}
		\eta_{\rm trans} = 0.75\left(1+\frac{\varv_{\rm esc}^2}{\varv_\infty^2}\right)^{-1}
\end{equation}
the wind is predicted to become optically thick. The pre-factor 0.75 reflects the calibration for very massive stars from \cite{Sabhahit_2023}. In practice, $\varv_\infty$ is approximated in the same study as
\begin{equation}\label{eq:v_term}
    \varv_\infty = V_{\rm e}^\infty \sqrt{\frac{2GM(1-\Gamma_{\rm e})}{R}}\left(\frac{Z}{Z_\odot}\right)^{0.2}
\end{equation}

Here, $V_{\rm e}^\infty$ represents the proportionality constant between the terminal velocity and the \textit{effective} escape velocity ($\varv_{\rm esc, eff}=\varv_{\rm esc}\sqrt{1-\Gamma_{\rm e}}$), rather than the standard Newtonian escape velocity. This distinction is crucial for massive stars approaching the Eddington limit, where the ($1-\Gamma_{\rm e}$) term significantly reduces the effective potential well. In the \cite{Sabhahit_2023} model, $V_{\rm e}^\infty$ is treated as a step function: it is fixed at 2.6 for hot stars ($T_{\rm eff}$\,$\geq$\,25~kK) and drops abruptly to 1.3 at lower effective temperatures. This approximation assumes a fixed wind velocity structure that changes abruptly only at the bistability jump, lowering both $\varv_\infty$, $\eta$, and consequently the transition threshold $\eta_{\rm trans}$ in cooler regimes. 
However, an alternative approach is to directly adopt tabulated $\varv_\infty$ profiles from detailed atmosphere models,  rather than a single analytical prescription \citep[e.g.][]{Krticka_2017,Bjorklund_2021,GormazMatamala_2023,Krticka_2025}.
Solving these equations, we show the existence of a natural correlation between the free-electron scattering $\Gamma_{\rm e}$-based transition and the multi-scattering one:

\begin{equation}\label{eq:actual_eta_trans}
    \eta_{\rm trans} = 0.75 \left( 1 + \frac{1}{(V_{\rm e}^\infty)^2(Z/Z_\odot)^{0.4}} \frac{1}{1-\Gamma_{\rm e}} \right)^{-1}
\end{equation}

\paragraph{Model uncertainties: $\eta$} 
The transition to thick winds is dominated by two compounding factors: the sensitivity of the wind efficiency parameter ($\eta$) to input physics and the structural assumptions used to calibrate the transition threshold ($\eta_{\rm trans}$), the calibration of the transition value $\eta_{\rm trans}$ is based on extensive studies of the Arches and 30~Doradus clusters \citep{Vink_2012,Sabhahit_2022}.
It is important to note that the transition mass-loss rate itself, derived empirically from observed transition luminosities, remains, at least within the calibration parameter space, phenomenologically robust. However, within a predictive 1D evolutionary framework, $\eta$ explicitly depends on the adopted mass-loss prescription (Equation~\ref{eq:eta}), representing the largest source of uncertainty due to its linear dependence on $\dot M$ and the wide mass-loss variability (see Section~\ref{subsec:thick_winds}). Stronger mass loss yields a faster increase in $\eta$, while a weaker one implies a delayed transition to the \gls{WR} phase (\citealt{GormazMatamala_2025}, \AtlasII).

\paragraph{Model uncertainties: $\eta_{\rm trans}$} 

In the most recent formulation of \cite{Sabhahit_2023}, the quantities parameterizing the transition to thick winds ($\Gamma_e$, $\eta$, and $\eta_{\rm trans}$) are not calculated using the star’s current parameters ($M$, $L$, and $T_{\rm eff}$), but are instead calculated over a fixed luminosity range ($10^5$ to $10^8 L_\odot$) to determine a hypothetical crossing point using the properties of a chemically homogeneous star derived from the fits of \cite{Grafener_2011}. This procedure appears designed to facilitate the implementation of the \cite{Vink_2011} mass-loss recipe, which adopts a "multiplier" to the optically thin mass-loss rates at the onset of thick winds. Because some very massive stars are already in the \gls{WR} phase near \gls{ZAMS}, they lack a physical transition point from thin to thick winds in their evolutionary history; therefore, the \cite{Sabhahit_2023} model effectively extrapolates a hypothetical pre-\gls{ZAMS} evolution to locate a theoretical anchor point for the \cite{Vink_2011} multiplier. 
This reliance on the chemical homogeneous mass introduces a systematic physical inconsistency. By fixing the transition calculation to chemical homoegeneous mass (i.e., the maximum possible mass for a given luminosity and surface-H abundance) the model adopts a mass consistently higher than that of an evolving star. This choice artificially lowers the Eddington factor ($\Gamma_{\rm e} \propto L/M$), thereby increasing the term $(1 - \Gamma_{\rm e})$. Following Equation~\ref{eq:actual_eta_trans}, a higher $(1 - \Gamma_{\rm e})$ leads to a higher value for the transition threshold $\eta_{\rm trans}$ (see also Figure A.1 from \citealt{Boco_2025}, in which a further analysis of the transition to thick winds is provided). \cite{Sabhahit_2023} adopt this approach because very massive stars above roughly 160~$M_\odot$ are thought to remain chemically homogeneous throughout the \gls{MS}. However, as shown by \cite{Hirschi_2025}, this assumption is not universally valid and is strongly metallicity- and model-dependent: at $Z = 0.002$, even stars approaching $300~M_{\odot}$ may not necessarily evolve under conditions of chemical homogeneity. Although \cite{Sabhahit_2023} acknowledge their framework as a "theoretical exercise" for high-redshift conditions, their reliance on a homogeneous approximation to preserve the validity of \cite{Vink_2001,Vink_2011} neglects actual surface properties, which is not instead the case for \cite{Sabhahit_2022}. This risks introducing significant errors in \gls{WR} phase duration and remnant mass, particularly for non-very massive, solar-metallicity, or low-metallicity stars where the homogeneity assumption breaks down.



\subsubsection{New multi-scattering model} 
\label{subsubsec:new_eta}

While the reliance on the chemically homogeneous approximation described in Section~\ref{subsubsec:multi_scatter} limits the general applicability of the \cite{Sabhahit_2023} framework, its underlying physical premise—that the transition to optically thick winds is governed by the wind efficiency $\eta$—remains robust. To preserve this physical consistency while moving beyond the calibration choices originally made to accommodate the \cite{Vink_2001,Vink_2011} models, we generalize the method to rely on the star's instantaneous surface properties. In order to calculate $\eta$ and $\eta_{\rm trans}$ more self-consistently without adhering to the original restriction, we recommend the implementation of the following updates to the canonical version:

\begin{enumerate}
    \item \textit{Temperature-dependent velocity scaling:}
    The behavior of $V_{\rm e}^\infty$ as a function of effective temperature remains a subject of active debate. While standard recipes enforce a sharp discontinuity, the work of \cite{Crowther_2006} and recent observations of low-metallicity B-supergiants suggest a complex picture. \cite{BerniniPeron_2024} report a drop in $V_{\rm e}^\infty$ occurring at $\sim$19~kK, significantly cooler than the canonical prediction at 25~kK. However, as noticeable in Figure~18 of \cite{BerniniPeron_2024}, the observational data exhibit considerable scatter. While $V_{\rm e}^\infty$ is statistically larger above 20~kK than below, the distribution of individual measurements does not uniquely demand a step-function jump. A continuous linear increase with temperature offers an equally plausible representation of the data within the errors and avoids enforcing a sharp transition at an uncertain temperature, as also supported by \cite{Verhamme_2024,Verhamme_2026}. We fit the $V_{\rm e}^\infty$ factor directly to the empirical observations of Galactic supergiants from \cite{Crowther_2006}. Figure~\ref{fig:vratio} shows the linear dependency from our best-fit model $V_{\rm e}^\infty = 1.59\times10^{-4}T_{\rm eff} - 0.89$. For $T_{\rm eff}\gtrsim$15~kK, our model yields a continuous range of values for $V_{\rm e}^\infty$ that generally exceed the fixed step-function values used in \cite{Sabhahit_2023}. Alternatively, rather than calculating $\varv_\infty$ with Equation~\ref{eq:v_term} and $V_{\rm e}^\infty$, a direct calculation is possible from the theoretical fits of, e.g., \cite{  Krticka_2025}, or the observational ones from \cite{Hawcroft_2024} and \cite{Alkousa_2026}.

\begin{figure}[!h]
\centering
\includegraphics[width=0.49\textwidth]{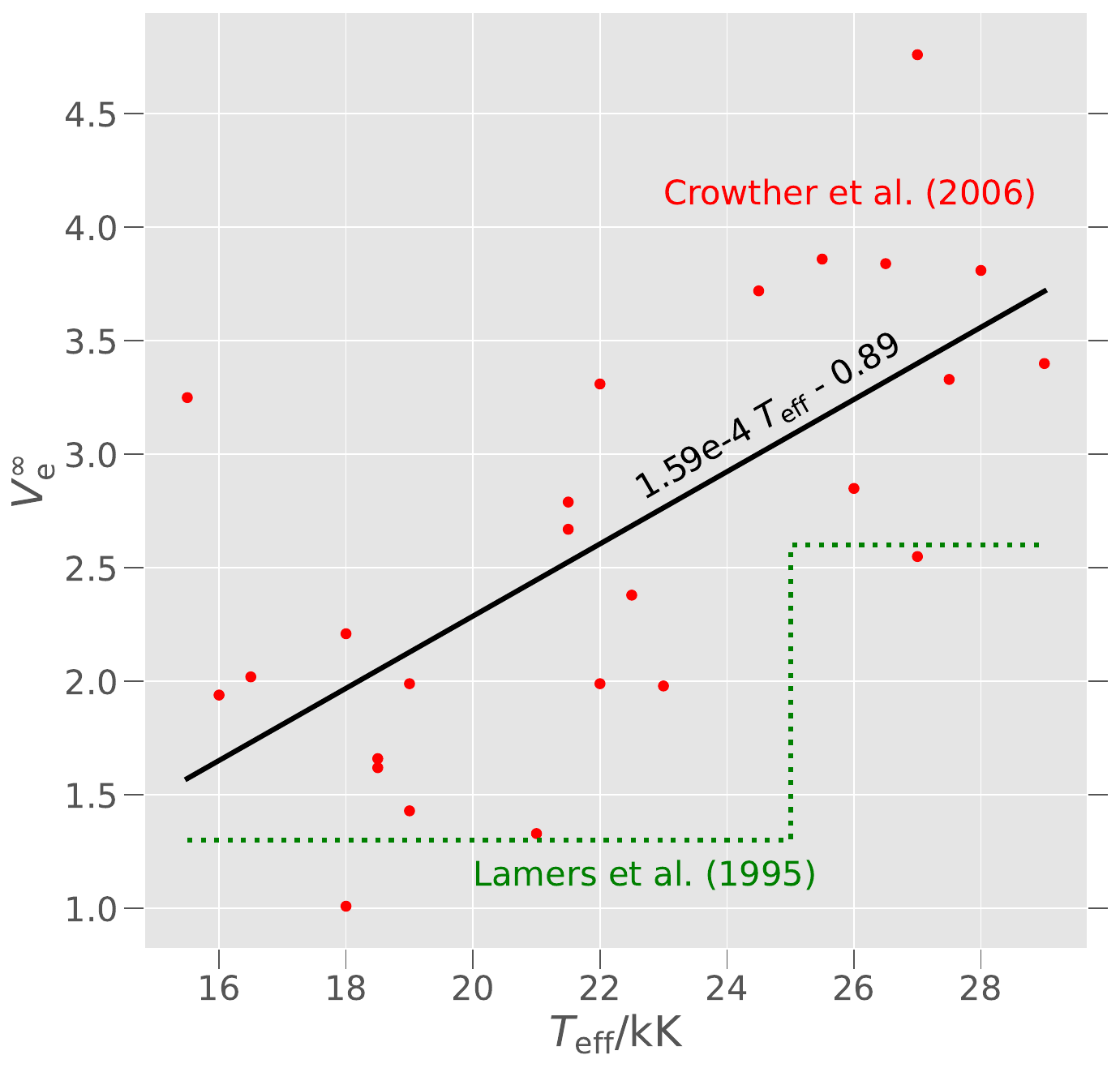}
\caption{$V_{\rm e}^\infty$ observations from \cite{Crowther_2006} as a function of effective temperature (dots). We also show our linear fit (black line) and the parametrization (green dotted line) from \cite{Lamers_1995}, used in many mass-loss models \citep[e.g.][]{Vink_2001,Vink_2021,Sabhahit_2023}.}
\label{fig:vratio}
\end{figure}

    \item \textit{Self-consistent $\varv_\infty$ and $\varv_{\rm esc}$ values:} Rather than calculating $\varv_\infty$ and $\varv_{\rm esc}$ from hypothetical chemically homogeneous conditions, the star's \textit{current} mass, as it was originally implemented in \cite{Sabhahit_2022}, luminosity, effective temperature and radius should be used in the calculations. 


    \item \textit{Thomson scattering $\Gamma{\rm_e}$ validity regime:} The transition to optically thick winds should not be allowed at $T_{\rm eff}\leq$15~kK, since this represents the limit of the extrapolation of the free-electron scattering physics behind the calculation of $\Gamma{\rm_e}$, upon which the entire calculation of $\eta$ and $\eta_{\rm trans}$ depends (see Section~\ref{subsubsec:coolWR} for more details). This also comes as a practical necessity because the terminal velocity fits can hardly be extrapolated beyond their original regimes. Figure~\ref{fig:v_term_fits} shows the terminal velocity as function of the star's effective temperature. These results show that any of the considered formulations heavily underestimates $\varv_\infty$ for the cooler side of line-driven winds, unless one simplifies this relation to the \cite{Lamers_1995} step function.

\begin{figure}[!h]
\centering
\includegraphics[width=0.49\textwidth]{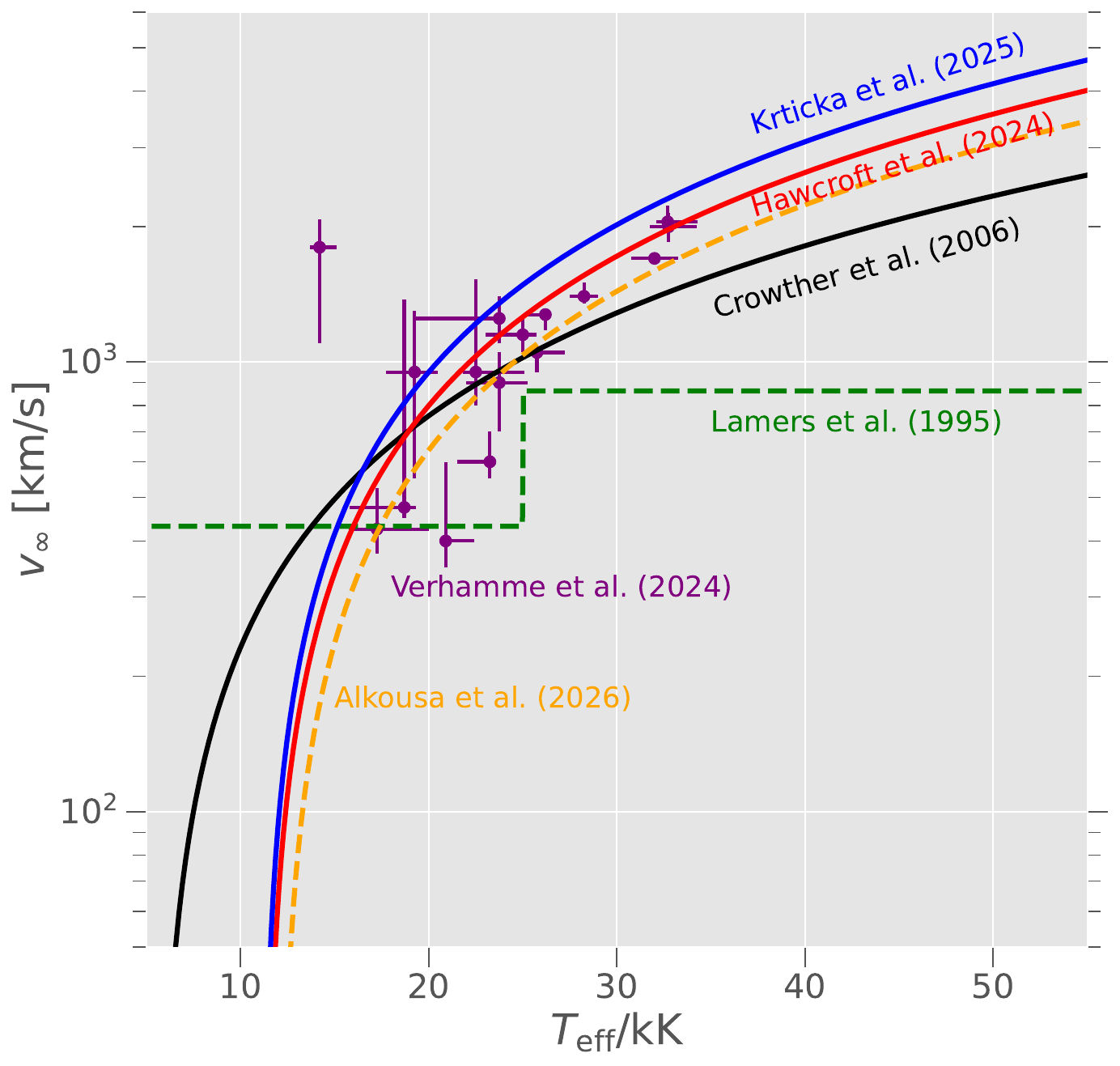}
\caption{Terminal velocity calculations for a 35~$M_\odot$ star at $Z_\odot$, $X_{\rm surf}\,=\,0.6$, and $\log(L/L_\odot)\,=\,5.4$ as a function of effective temperature. As a comparison, the observational data from \cite{Verhamme_2024} is also shown. Beyond the observed parameter space, every model leads to a drastic reduction for $\varv_\infty$ at $T_{\rm eff}\lesssim$\,10~kK.}
\label{fig:v_term_fits}
\end{figure}

    \item \textit{Proximity to the Eddington limit:}
    Very massive stars possess surfaces that evolve close to the Eddington limit. Relying solely on the $\eta$ parameter to trigger thick winds can result in an unphysical delay in the onset of the \gls{WR} phase, particularly when using modern optically thin wind recipes. Because these newer prescriptions often predict lower mass-loss rates than the \cite{Vink_2001} baseline used to calibrate $\eta$ and $\eta_{\rm trans}$ in \cite{Sabhahit_2023}, the resulting $\eta$ values may remain artificially low. We therefore recommend supplementing the $\eta$ criterion with a $\Gamma_{\rm e} > \Gamma_{\rm e, trans}$ ($\Gamma_{\rm e} > 0.5$ for $Z\geq0.008$; \citealt{GormazMatamala_2022b}) condition to ensure that the physical effects of high Eddington factors are explicitly captured.

\end{enumerate}

We emphasize that all transition criteria discussed in this Atlas, whether based on $\Gamma_{\rm e}$, $\eta$, or surface abundance thresholds, are subject to the inherent approximations of 1D hydrostatic modeling. No single criterion for such models is derived from pure first-principles physics; instead, they serve as calibrated stellar engineering solutions intended to approximate the onset of dense outflows. Although the wind efficiency parameter $\eta$ can respond more dynamically to instantaneous surface conditions, its application still relies heavily on the underlying parameterizations.

\subsubsection{Optical depth approximation transition}
\label{subsubsec:tau_transition}

A different physical criterion for the transition to optically thick winds evaluates the optical depth parameter of the wind itself \citep{AguileraDena_2022,Pauli_2022b}. Unlike transition conditions based on the surface hydrogen abundance $X_{\rm surf}$ or the Eddington factor $\Gamma_{\rm e}$, the parameter $\tau_{\rm wind}$ directly tracks how much the stellar radiation is trapped by the outflowing material, following the equation below:

\begin{equation}
    \tau_{\rm wind} = \frac{\kappa_{\rm e}|\dot M|}{4\pi R(\varv_\infty - \varv_0)}\log\left(\frac{\varv_\infty}{\varv_0}\right)
\end{equation}

Where $\varv_0$ is the expansion velocity near the stellar surface taken as 20~$km\,s^{-1}$.
The calculation relies on integrating the continuity equation alongside a prescribed velocity law \citep{deLoore_1982,Langer_1989b}.

For the formation of hydrogen-rich \gls{WR} stars, recent studies propose a threshold value of $\tau_{\rm wind} \approx 0.2$ \citep{Sen_2023}. We note that this approach represents the least adopted transition model in the current explored literature. Most evolutionary codes do not calculate $\tau_{\rm wind}$ directly.

\subsubsection{Mass-loss rates}

\begin{figure}[!ht]
\centering
\includegraphics[width=0.499\textwidth]{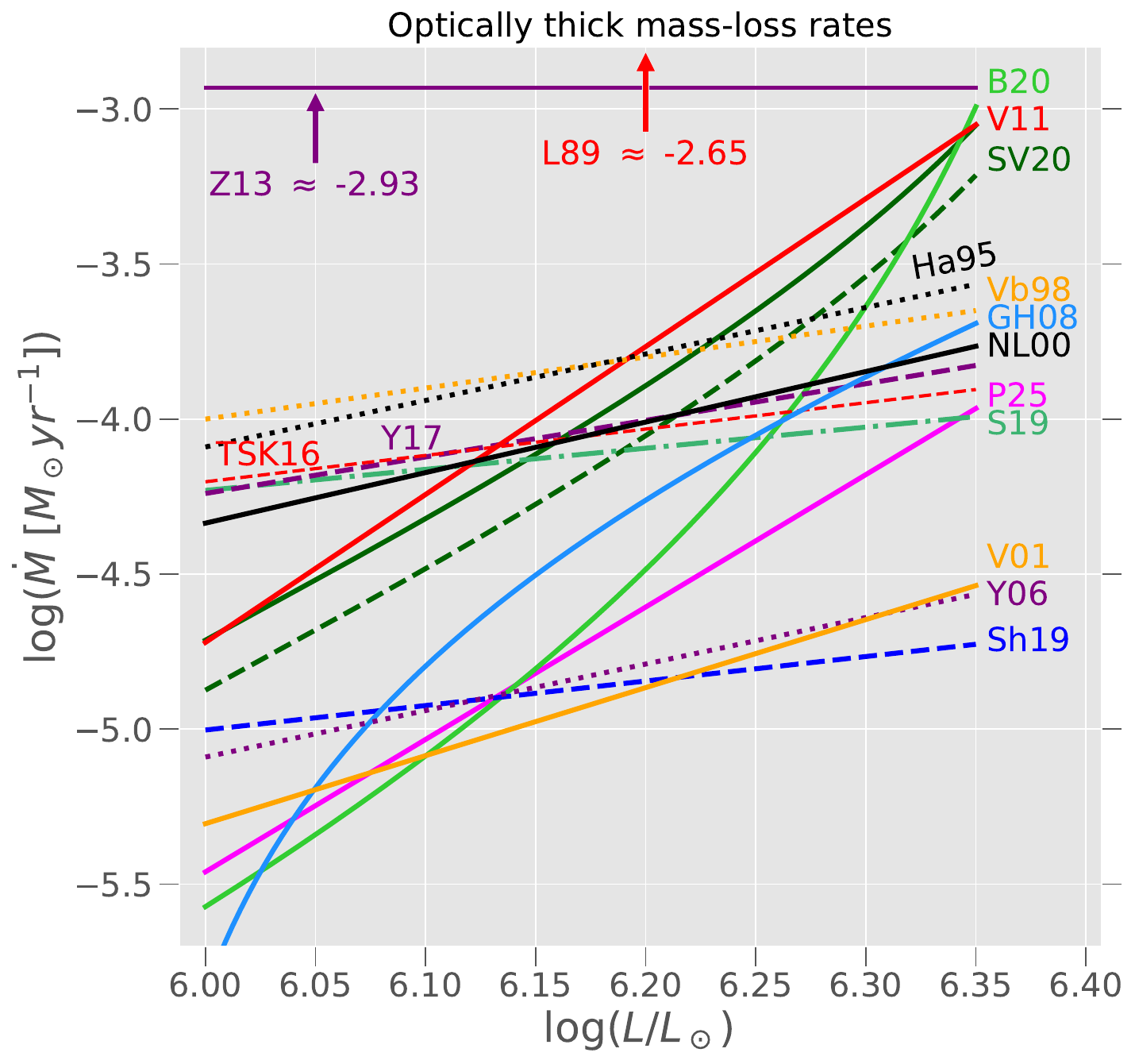}
\caption{Evolution of mass-loss rates for optically thick winds as a function of luminosity for a 60~$M_\odot$ star at $Z_\odot$, surface hydrogen fraction $X_{\rm surf}\,=$\,0.4, and $T_{\rm eff}$\,=\,30~kK, assuming $\Gamma_{\rm e}$\,$\geq$\,0.4. 
For comparison, the pre-bistability mass-loss rates from \cite{Vink_2001} are shown.
The dark green dashed green dark line shows the \cite{Sander_2023} temperature correction applied to the \cite{Sander_2020} mass loss for $T_{\rm eff}=$150~kK.}
\label{fig:thick}
\end{figure}

Figure~\ref{fig:thick} shows the mass-loss rates predicted by several  commonly used recipes for optically thick winds, evaluated for a 60~$M_\odot$ star at $X_{\rm surf}$\,=\,0.4 (a common transition point to optically thick winds; \citealt{Glebbeek_2009,Ekstrom_2012}) and $T_{\rm eff}$\,=\,30~kK, a threshold often adopted in stellar evolution models for fully-ionized surfaces \citep{Josiek_2024,Romagnolo_2024,GormazMatamala_2025}. 

The historical prescription from \cite{Langer_1989} predicts mass-loss rates significantly higher than modern estimates, appearing as an upper bound in Figure~\ref{fig:thick} ($\log \dot{M} \approx -2.65$). This disparity arises because L89 was explicitly calibrated to match the studies of that era \citep[e.g., ][]{van_der_Hucht_1986,de_Freitas_Pacheco_1988,Schmutz_1989} which argued that previous standard rates \citep[][]{Barlow_1981,Abbott_1986} were underestimated by factors of 2--3. L89 consequently adopted a steep mass-dependence ($\dot{M} \propto M_\text{WR}^{2.5}$) to reproduce these higher values. This also arises from extrapolation, since we are applying a mass-loss formula that was fit for a sample of \gls{WR} stars of at most 25~$M_\odot$ to a 60~$M_\odot$ star. Similarly, \cite{Zdziarski_2013} developed their mass-loss formulation from the \cite{NugisLamers_2000} WN sample corrected for clumping, with a maximum mass of $M_{\rm WN}\approx22~M_\odot$. As also highlighted by the same authors, extrapolating their mass-loss rates beyond such a mass results in considerable overestimations.

We also show that the mass-loss rates from \cite{Vink_2001}, despite representing the weaker optically thin wind regime, predict similar mass-loss rates to \cite{Yoon_2006} and \cite{Shenar_2019} at $\log L \gtrsim6.3$, and higher than \cite{Grafener_2008,Bestenlehner_2020,Pauli_2025} for $\log L \lesssim 6.05$. As highlighted by \cite{GormazMatamala_2022}, the low mass-loss rates from \cite{Bestenlehner_2020} are due to the fact that the star has not yet reached the transition condition to optically thick winds at $\Gamma_{\rm e}\,\approx\,$0.5, which is a requirement for their application. On the other hand, the underestimate of mass loss from \cite{Grafener_2008} is a well known issue in stellar evolution, which leads, for instance, the {\tt GENEC} models within this regime to select at each timestep the maximum mass loss between \cite{Grafener_2008} and \cite{Vink_2001}, rather than only applying one wind recipe at a time \citep{Eggenberger_2021,Hirschi_2025} (more details in Section~\ref{sec:other_uncert}).

Comparing these formulations numerically nevertheless only reveals part of the discrepancies. The fundamental reliability of these rates and the transitions between them hinges critically on the physical parameters governing the wind driving mechanism. Central to this issue is the application of the Eddington factor $\Gamma_{\rm e}$, which, as we discuss next, faces significant physical contradictions when applied to cooler stars.



\subsubsection{Cool WR problem: $T_{\rm eff}$-dependency of $\Gamma_{\rm e}$}
\label{subsubsec:coolWR}

A central theoretical bottleneck in modeling evolved massive stars is the `cool Wolf-Rayet problem', which is a regime where the standard physical driver of mass loss loses its consistency and detections are too sparse for calibrating stellar wind models. 

The winds of massive stars are deeply dependent on the relationship between luminosity and mass, a ratio commonly represented by $\Gamma_{\rm e}$. Mass-loss rates for both optically thin \citep[e.g.][]{Pauli_2025} and optically thick winds \citep[e.g.][]{Bestenlehner_2020} explicitly depend on $\Gamma_{\rm e}$ in their formulation. Furthermore, the most widely used criteria for the transition from optically thin to optically thick winds are critically linked to this factor. For the free-electron scattering condition, it is assumed that a star develops optically thick winds when $\Gamma_{\rm e}$ exceeds a specific calibrated threshold. Similarly, for the multiple-scattering condition, while the transition is formally determined by the wind efficiency parameter $\eta$, in its original formulation the onset of the strong \gls{WR} winds is also tied to an increasing $\Gamma_{\rm e}$ \citep{Sabhahit_2022,Sabhahit_2023}.

Models adopting $\Gamma_{\rm e}$ to stars outside a very specific, high-temperature regime introduce a significant physical inconsistency. The derivation of $\Gamma_{\rm e}$ is fundamentally based on the assumption that the stellar opacity is constant and dominated by a single process: Thomson scattering by free electrons in a fully ionized plasma \citep{Eddington_1926}. This "continuum opacity" assumption has a strict domain of validity. The physical prerequisite for electron scattering to be the dominant opacity source is a stellar photosphere that is almost completely ionized, in particular for its approximated formulation of Equation~\ref{eq:logGammaEdd}. This condition is only met in the hottest O-type and very early B-type stars, which have effective temperatures $T_{\rm eff}\,\gtrsim\,$30~kK \citep{Bestenlehner_2020}. While 30~kK represents a robust threshold for absolute Thomson dominance where H, He, and CNO elements are fully ionized, a transition regime of functional validity may persist down to approximately $15$ kK. In this range (15~kK\,<\,$T_{\rm eff}$\,<\,30~kK), electron scattering can still play a role in the opacity levels \citep{BerniniPeron_2025} due to the potentially high ionization levels \citep[see also the Saha ionization equation:][]{Saha_1920,Saha_1921}.
However, a hard physical floor exists at $\approx 15$~kK. Below this temperature, the recombination of hydrogen introduces an `opacity cliff' \citep{Davidson_1987,Owocki_2015}. As a star's photosphere cools, it is no longer fully ionized and the true Rosseland mean opacity increases, becoming orders of magnitude larger than the constant electron scattering value used to calculate the classical $\Gamma_e$. Consequently, any model that uses the classical $\Gamma_e$ for stars cooler than 30 kK is not correctly estimating the true radiative force acting on the stellar material. This physical mismatch is the root of the "cool WR problem" and represents a fundamental flaw in several current mass-loss models.

The necessity of treating ionization rigorously is reinforced by X-ray observations of massive binaries. \cite{Lai_2024} demonstrate that neutral wind models fail to reproduce key observational features in the color-color diagrams of Cyg~X-1, explicitly requiring a partially ionized absorbing medium. This empirical evidence shows that assuming constant ionization states (like fully ionized Thomson scattering) is insufficient even for O-supergiants, let alone the cooler WR regimes discussed here.

It must be however higlighted that, despite not being within its regime of validity, $\Gamma_{\rm e}$ may still be used as a 0th-order approximation for the $L/M$ ratio. Despite not capturing the actual conditions on stellar surfaces, it still acts as a qualitative indicator for near-Eddington winds, regardless of its extrapolation uncertainties.

\paragraph{Mass loss} At present, no dedicated mass-loss recipe exists for cool \gls{WR} stars. Recent modelling of late-type WNh stars from \cite{Lefever_2025} directly confronts this cool \gls{WR} problem and illustrates why simple solutions are inadequate. Their work reveals that, far from a smooth extrapolation, the mass loss in this regime is complex and highly sensitive to effective temperature and metallicity.
They confirm that applying prescriptions based on the classical $\Gamma_{\rm e}$ is flawed, not just because the opacity source is wrong (it is highly dependent on Fe opacity gradients), but because the underlying physics leads to non-monotonic and even discontinuous behavior. This forces current evolutionary codes to employ ad-hoc workarounds. Some models revert to optically thin wind recipes for cool \gls{WR} stars, even when the classical $\Gamma_{\rm e}$ exceeds the transition threshold (e.g., \citealt{Romagnolo_2024,GormazMatamala_2025}, \AtlasII).
Others, such as the canonical Dutch winds or models using either \cite{Grafener_2008} or \cite{Sabhahit_2023}, continue to apply strong, optically thick winds. While applying strong winds to these objects may better align with observational constraints like the \gls{HD} limit (\citealt{Boco_2025},\AtlasII)
, it is important to recognize that the theoretical justification based on the classical $\Gamma_{\rm e}$ is physically unsound in this temperature regime, with recent studies suggesting weaker mass-loss rates \citep{BerniniPeron_2025,Lefever_2025}.

\paragraph{Transition to thick winds} 
The reliance on an inconsistent physical parameter is particularly problematic for the transition criteria themselves. Both the free-electron scattering and the \cite{Sabhahit_2023} multiple-scattering transition models have limitations: they check for the initiation of optically thick and/or strong \gls{WR} winds as a function of the classical $\Gamma_{\rm e}$, even when the star cools below 30~kK. This methodology weakens the robustness of both prescriptions. It forces the models to rely on uncertain extrapolations rather than on additional recalibrations that attempt to empirically account for the missing physics of atomic opacities. A truly predictive model for the onset of optically thick winds must incorporate a more sophisticated treatment of opacity that is valid across the full range of temperatures encountered by evolving massive stars.

\subsection{Variability with $Z_\odot$ calibrations}
\label{subsec:calibrations}

\begin{figure*}
\centering
\includegraphics[width=0.75\textwidth]{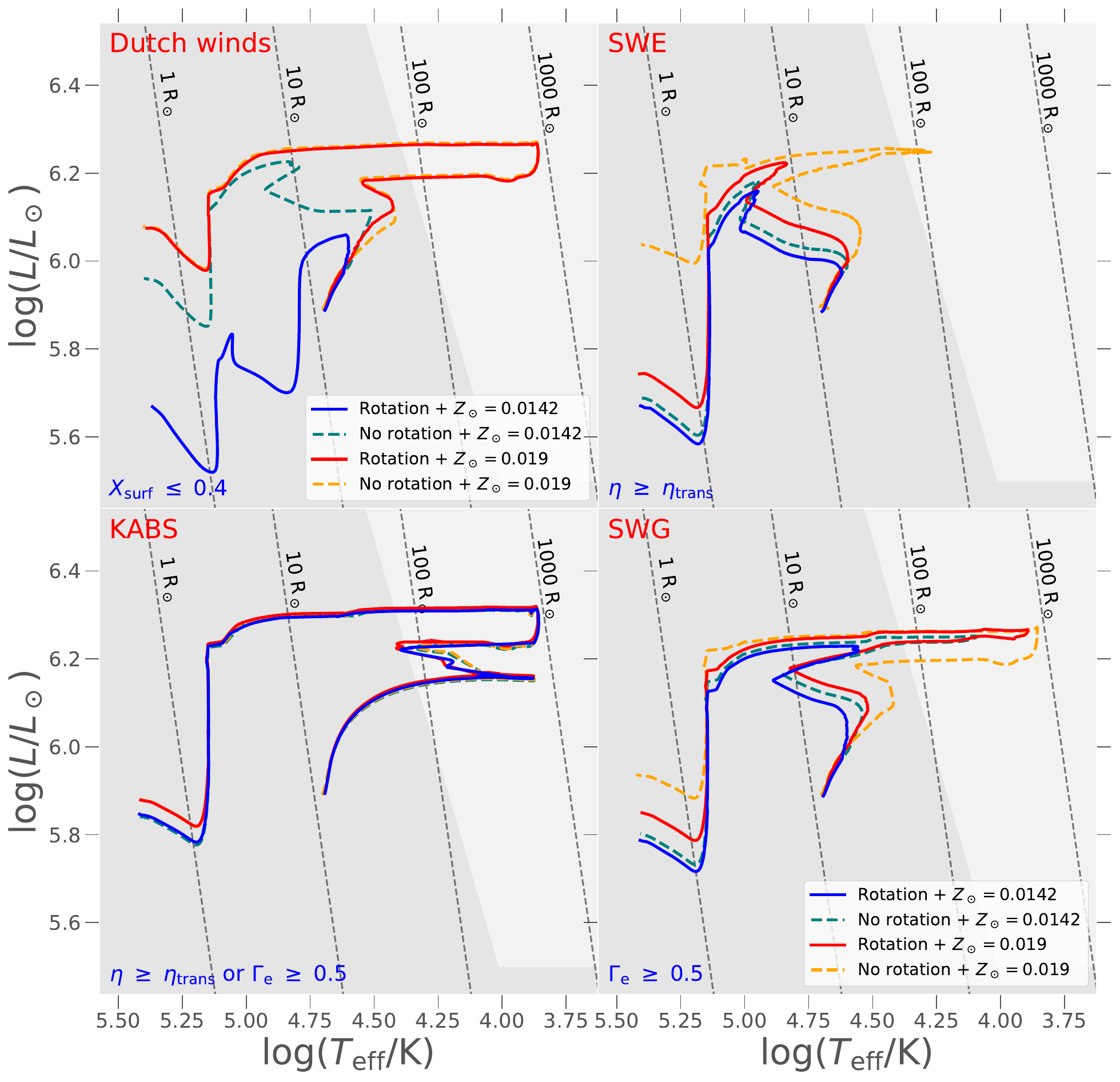}
\caption{\gls{HR} diagram for the evolution of 75~$M_\odot$ stars at $Z=0.0142$ until the depletion of carbon in the core for four different stellar winds recipes and different transition conditions to the \gls{WR} phase (see Section~\ref{subsec:thick_winds}). Each panel shows stars with $\Omega$/$\Omega_{\rm crit}=0.4$ at \gls{ZAMS} (solid lines) and  non-rotating stars (dashed lines)  for two values for $Z_\odot$.  The \gls{LBV} region beyond the \gls{HD} limit is indicated (white). A higher $Z_\odot$ value leads to lower mass-loss rates and to a wider expansion. With rotation, the star enters the \gls{WR} phase earlier in its evolution.
}
\label{fig:75_HR}
\end{figure*}

Wind formulations usually scale mass loss with metallicity as $\dot M \propto (Z/Z_\odot)^\alpha$, with $\alpha$ a set calibration factor and $Z_{\odot}$ representing solar metallicity. Since solar metallicity does not have an universally recognized value (0.019: \citealt{Anders_1989}; 0.017: \citealt{Grevesse_1996}; 0.0142: \citealt{Asplund_2009}), the choice of the $Z_\odot$ scaling factor can vary across different studies. For instance, $Z_\odot$ was set to 0.019 in the \cite{Vink_2001} original fits and to 0.0142 in \cite{Krticka_2024}, while for \cite{Pauli_2025} the metallicity-dependency was not calibrated as a function of $Z_\odot$, but only as a function of the Fe abundance. All these calibrations are meant to represent the strength of the Fe lines in winds, but when different formulations are combined within stellar models to simulate different evolutionary phases of a given star (e.g. thin and thick winds), the $Z_\odot$ scaling factor may have different calibrations for each adopted mass-loss formula. Under these conditions, common modeling choices are to either adopt the original $Z_\odot$ values from each mass-loss prescription \citep[as is the case for e.g.,][]{Boco_2025,Costa_2025,Romagnolo_2025}, or to merge all the different metallicity dependencies under one single $Z_\odot$ value (e.g., some {\tt GENEC} and {\tt POSYDON} models, such as \citealt{Eggenberger_2021,Bavera_2023,Kruckow_2024,GormazMatamala_2025,Hirschi_2025}, or \AtlasII). 

While adhering to original calibrations preserves the specific original fits, standardizing $Z_\odot$ facilitates rigorous comparisons across mass-loss rates and offers better physical self-consistency by aligning the wind-driving Fe-group opacities with the modeled metal distribution governing stellar interiors via opacity tables. One choice is more consistent with the original wind studies while the other ensures greater physical self-consistency within evolutionary codes. Ultimately, both approaches are not universally valid and fall out of self-consistency once opacity and line-driven winds are extrapolated outside the Milky Way. This is due to the fact that Fe does not scale linearly with oxygen (i.e., a proxy of a star's total metallicity), with only $\sim$30\% of stars with near-solar relative abundances in the local universe, and nearly none in higher redshift environments \citep{Chruslinska_2024, Chruslinska_2025}.
Since both approaches possess limitations and validity depending on the goals of a study, we recommend that future research explicitly specifies which of the two paths is taken to ensure reproducibility.

To illustrate the effect of choosing different $Z_{\odot}$ calibrations, Figure~\ref{fig:75_HR} shows the \gls{HR} evolution of a 75~$M_\odot$ star at $Z=0.0142$ combining rotation and the application of either $Z_{\odot}=0.019$ or $0.0142$ for the four different models (Section~\ref{subsubsec:wind_implem}). 
Evolutionary models that use inherently strong \cite{Vink_2001} thin winds exhibit a high degree of sensitivity to the chosen $Z_{\odot}$ value. This variability arises because the $Z_{\odot}$ calibration and the inclusion of rotation act as multiplicative effects on the already high mass-loss rates, heavily impacting the star's hydrogen envelope depletion. Consequently, these strong wind models show a more robust expansion when using a higher $Z_{\odot}$ (i.e., weaker mass loss) for calibration. On the other hand, weaker thin winds like \cite{Krticka_2024}, shown in the bottom-left panel, do not display comparable changes in evolutionary outcomes across the same parameter space.

Finally, we caution that the standard power-law scalings ($\dot M \propto Z^\alpha$) implicitly assume that the wind driving mechanism remains dominated by the same set of lines (primarily Fe) across all metallicities. Recent radiation-hydrodynamic studies suggest this assumption breaks down at low metallicity \citep[$Z\leq 0.1~Z_\odot$;][]{Krticka_2025}, where the dominant opacity source shifts from Fe to CNO elements. This transition introduces non-linearities in the \textit{Z}-dependence that simple scaling laws cannot capture. We show its implications for mass-loss uncertainties in Section~\ref{sec:other_uncert}.


\section{Cool Supergiant winds}
\label{sec:dust_driven}
In isolated conditions, non-very massive stars possess insufficient luminosity to drive strong, optically thick winds during the \gls{MS}. To evolve into \gls{WR} stars, these objects require efficient internal mixing and strong mass loss during the cool \gls{YSG} and \gls{RSG} phases \citep{Georgy_2012}. Current prescriptions for these phases, however, remain highly uncertain, with rates varying by orders of magnitude across different models \citep{Antoniadis_2024}.

Mass loss for \glspl{YSG} (4~kK\,$\lesssim$\,$T_{\rm eff}$\,$\lesssim$\,8~kK), also known as hypergiants, is highly uncertain due to few observations within this so-called ``yellow void" \citep{Nieuwenhuijzen_deJager_1990}. Additionally, both pulsations and outbursts, which cannot be self-consistently simulated in 1D, may play a more important role than continuous winds \citep{Koumpia_2020,Humphreys_2023}. On the other hand, \gls{RSG}s ($T_{\rm eff}$\,$\lesssim$\,4~kK) are more frequently detected and studied. Given the lack of dedicated models for the \gls{YSG} phase, it is common practice to adopt for this regime the same type of mass loss that is applied to \gls{RSG}s. 

\begin{figure*}[!t]
\centering
\includegraphics[width=0.99\textwidth]{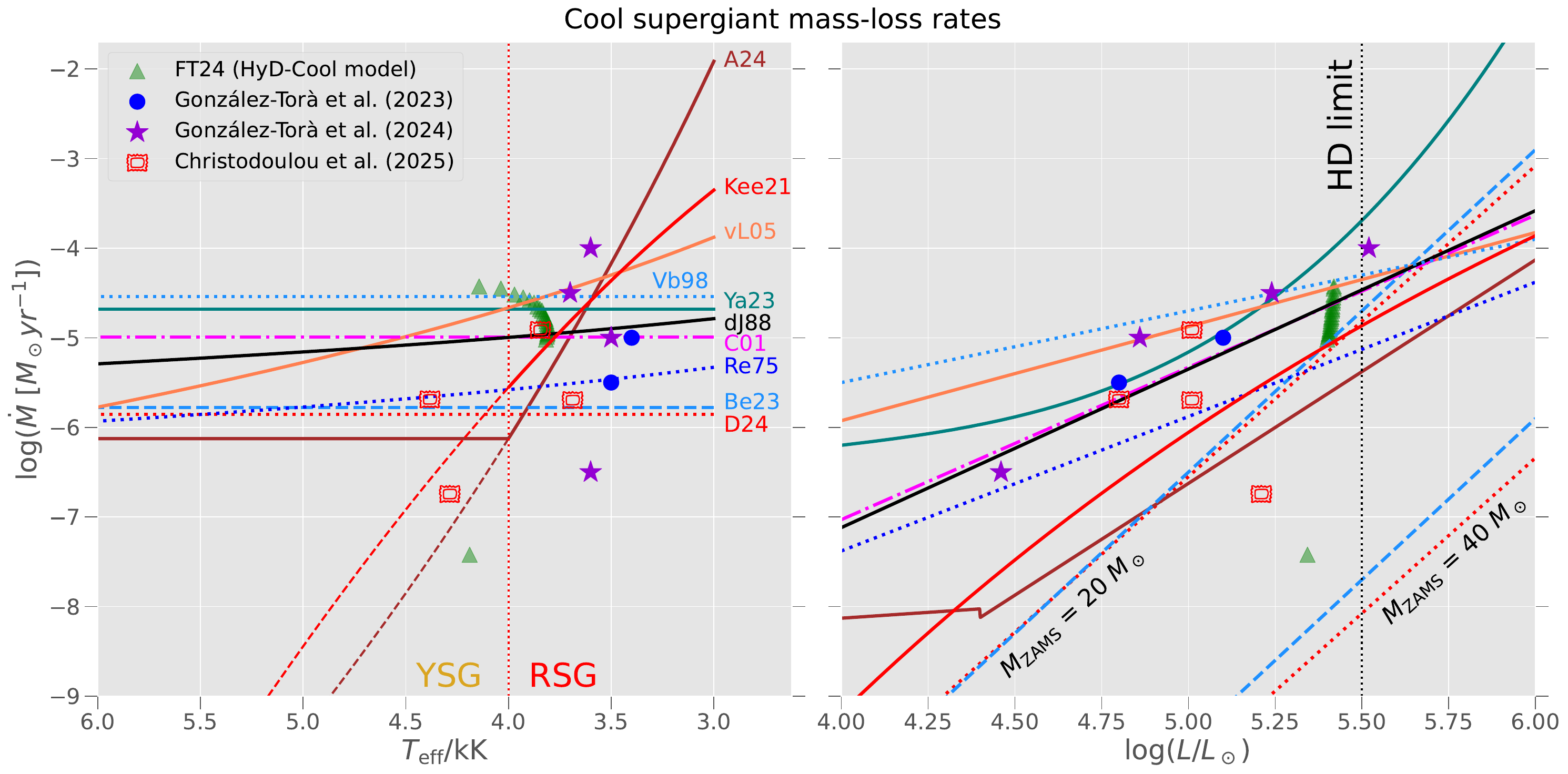}
\caption{
Mass-loss rates of cool supergiant winds as a function of effective temperature (left) and luminosity (right). 
For the left panel we adopt a constant $\log L = 5.2$, while for the right panel we assume a constant $T_{\rm eff} = 4$~kK. The initial mass was set to $20~M_\odot$ for the calculation of the mass-loss rates of \cite{Beasor_2023} and \cite{Decin_2024} in the left plot, and to both $20~M_\odot$ and $40\,M_\odot$ for the right plot calculations. \cite{Kee_2021}, with the dashed line representing its extrapolation for \glspl{YSG}, was calculated assuming a Rosseland mean opacity of 0.01~$cm^2/g$ and a turbulent velocity of 18.2~$km/s$. Blue circles, violet stars, and red ``ravioli'' symbols denote the observational measurements from \cite{GonzalezTora_2023}, \cite{GonzalezTora_2024}, and \cite{Christodoulou_2025}, respectively, as a reference. The \cite{Fuller_2024} mass-loss rates from the HyD-Cool model are shown as green triangles, for a temporal separation of 0.01~Myr each. 
The dashed red and brown lines show the extrapolation of \cite{Kee_2021} and \cite{Antoniadis_2024} mass loss extended beyond their original temperature range, illustrating the behavior if mass-loss rates were not truncated at $T_{\rm eff} = 4$~kK. 
In the left panel, the vertical red dotted line marks the approximate boundary between \gls{YSG} and \gls{RSG} regimes, while in the right panel the black dotted line indicates the \gls{HD} limit. 
For both panels, the cool-supergiant mass adopted for Re75 is fixed at $20~M_\odot$. NdJ90 is not shown since it nearly completely overlaps with dJ88.
}
\label{fig:dust}
\end{figure*}

The initiation temperature of cool supergiant winds differs substantially between studies. Some {\tt FRANEC} models initiate them at $T_{\rm eff}\leq$\,12~kK \citep{Ugolini_2025}, encompassing \gls{BSG}, \gls{YSG}, and \gls{RSG}s. In contrast, some {\tt MESA}-based models \citep[e.g.][]{Gilkis_2021,Romagnolo_2024,Liotine_2025} initiate them at $T_{\rm eff}\leq$\,10~kK, {\tt GENEC} at $\log T_{\rm eff}\leq$\,3.9 \citep{Eggenberger_2021}, {\tt COMPAS} at  $T_{\rm eff}\leq$\,8~kK \citep{Merritt_2025}, and others restrict them to only \gls{RSG}s \citep[e.g.][]{Vink_2021}.

Following the results of \cite{Zapartas_2025}, the $T_{\rm eff}$ initiation threshold for these winds can be altered without facing strong differences within the \gls{HG}. This is because this phase is fast enough for mass loss to not play a substantial role. However, once stars enter the slower \gls{CHeB} phase, whether cool supergiant winds are initiated or not will considerably impact their evolution.

For non-very massive stars, mass loss from optically thin winds alone is not strong enough to eject enough envelope mass to cause stars to become \gls{WR}. Canonically, the widely used \cite{deJager_1988} model prescribes considerably high \gls{Mdot} that can lead most \gls{BH} progenitors at $Z_\odot$ to evolve into \gls{WR} stars before their collapse \citep{Romagnolo_2024}. This effect could also be more evident with a higher or similar \gls{Mdot} \citep{vanLoon_2005,Goldman_2017,GonzalezTora_2023,Yang_2023,GonzalezTora_2024}, while it could be weakened at lower rates \citep[e.g.][]{Beasor_2021,Beasor_2023,Decin_2024,Antoniadis_2024,Antoniadis_2025}.

\subsection{Mass loss}

Mass-loss rates for cool supergiants may differ by orders of magnitude depending on which recipe is adopted, with the widest disparities arising for \glspl{YSG} and beyond the \gls{HD} limit.

Figure~\ref{fig:dust} shows a selected sample of cool supergiant mass-loss recipes, on top of samples based on interferometry data \citep{GonzalezTora_2023,GonzalezTora_2024} and observations of low-metallicity galaxies \citep{Christodoulou_2025}. The most recent studies show considerable variability in mass-loss rates, which can oscillate between comparable to the most widely used \cite{deJager_1988} ones \citep{vanbeveren_1998,Crowther_2001,GonzalezTora_2023,Yang_2023,GonzalezTora_2024}, to orders of magnitude lower \citep{Beasor_2023,Antoniadis_2024,Decin_2024}, with \cite{Beasor_2023} and \cite{Decin_2024} being the only two mass-loss recipes that match remarkably well between each other due to the analysis of the same supergiant sample by the respective authors. Additionally, the extrapolation to $T_{\rm eff}\,>\,4$~kK of the mass-loss rates from \cite{Antoniadis_2024} would lead to underestimate mass loss, but keeping instead the temperature dependency of mass loss past this threshold to a fixed $T_{\rm eff}\,=\,4$~kK shows a good agreement with what is prescribed by \cite{Beasor_2023} and \cite{Decin_2024}.

The various recipes for cool supergiant winds can be distinguished by their primary physical dependencies. The models from \cite{Beasor_2023} and \cite{Decin_2024}, for instance, are unique in that they are parametrized without an explicit effective temperature dependency. Instead, their mass-loss rates are a function of luminosity and $M_{\rm ZAMS}$. While mass loss in these models does increase with luminosity, this trend is misleading because it is coupled with a strong inverse dependency on $M_{\rm ZAMS}$, where a higher initial mass leads to considerably lower mass loss. This competing effect is so significant that in our models a star with a high initial mass of 75~$M_\odot$, i.e., well above the most massive \glspl{RSG} they used to fit their formulae ($\sim$25~$M_\odot$), is calculated to have an extremely low mass-loss rate ($\log(\dot M~[M_\odot yr^{-1}])\,\approx\,$-11, i.e. at least six orders of magnitude than \citealt{deJager_1988}), despite its high luminosity. However, given the intrinsic dependency of such prescription on $M_{\rm ZAMS}$, a cool supergiant forming from the merger of two stars in a binary may have considerably more mass and luminosity than an isolated star at the same $T_{\rm eff}$ \citep{Schneider_2024}, and might lead to mass-loss rates that are orders of magnitude higher for cool supergiant winds. More generally the factor $M_{\rm ZAMS}$ loses meaning once stars are not in isolation, and a good practice should be to model their mass loss as a function of its current parameters, and not as an explicit function of its history. 

In contrast, other models incorporate a direct temperature dependence, generally showing higher mass loss at cooler temperatures. {This inverse $T_{\rm eff}$ dependence can be expected because more evolved \glspl{RSG} become cooler and their photospheric temperatures are weakly anti-correlated with metallicity (due to opacity). This extreme steepness is a direct result of the $T_{\rm eff}$ scale employed: \cite{Antoniadis_2024} uses a $T_{\rm eff}$-relation based on the \cite{Tabernero_2018} sample calibrated on \textit{i}-band spectra, which yields a significantly narrower range of effective temperatures for \glspl{RSG} at a given metallicity compared to traditional TiO-band measurements. While these \textit{i}-band measurements are more reliable because they are less affected by mass-loss-induced atmospheric perturbations \citep{Davies_2013}, \cite{vanLoon_2025} cautions that such a steep dependence may be unrealistically strong and physically difficult to justify. Furthermore, the root of the discrepancy between the stronger and weaker mass-loss rates reported in the literature remains a subject of active debate between model-driven and population-driven perspectives.}

\cite{Antoniadis_2024} argue that the primary cause of these discrepancies lies in the divergent assumptions and methodologies utilized across different studies, such as variations in radiation transfer modeling and parameterization. However, \cite{vanLoon_2025} suggests that selection effects play a central role, particularly the bias toward sampling either short-lived, extreme mass-loss phases or more moderate, representative ones. These intense, short-lived superwind phases can exceed nuclear mass consumption rates by a factor of ten or more, yet because they are rare, it is not clear which regime dominates the total integrated mass loss over the \gls{RSG} lifetime. Consequently, the current lack of a singular consensus likely stems from a combination of these factors: the inherent uncertainties of disparate modeling techniques and the stochastic nature of the mass-loss events themselves.

\subsubsection{HD limit}
The application of cool supergiant winds beyond the \gls{HD} limit at $\log L\,>\,5.5$ \citep{Davies_2018,McDonald_2022} represents a modeling necessity that is nevertheless fully unconstrained from the absence of any \gls{YSG} and \gls{RSG} calibration points.
The extrapolation of such recipes leads mass-loss rates to quickly converge to values that are similar or higher than the legacy ones of \cite{deJager_1988}, due to their strong dependence on luminosity.
It is worth noting that while empirical power-law recipes may underestimate mass loss in this extreme regime, physical models based on convective shocks predict a "runaway" mass-loss phase for the most massive RSGs ($M_{\text{ZAMS}} \ge 30 M_{\odot}$) \citep{Fuller_2024}.  In this scenario, as the star expands and escape velocity drops, the boil-off rate increases exponentially, potentially stripping the hydrogen envelope much faster than predicted by canonical rates (e.g., \citealt{deJager_1988}). Additionally, runaway dynamical mass loss driven by strong shocks from growing pulsations offers a complementary pathway to efficiently strip most stars above a certain mass threshold, alleviating the missing RSG problem \citep{Bronner_2025,Suzuki_2025,Laplace_2026,Sengupta_2026}. Both of these physical mechanisms provide a potential solution to the scarcity of observed \glspl{RSG} above a luminosity limit of $\log(L/L_{\odot}) \approx 5.5$ by naturally shortening the \gls{RSG} lifetime of high-mass progenitors.
The potential for a mass-loss runaway in this regime may also stem from the decoupling of luminosity and current mass in \glspl{RSG} shown in \cite{Farrell_2020}. Building on these results, \cite{Vink_2023c} proposed that applying the hot-star V11-$\Gamma_{\rm e}$ boost to cool supergiants naturally strips the envelope, acting as an alternative or complementary mechanism to the convective boil-off proposed by \cite{Fuller_2024}. However, \cite{Fuller_2024} argue that this analogy to hot-star line-driven winds may be limited, as there is currently no established physical model for radiatively driven winds in cool stars via molecular absorption lines. They further contend that while radiation pressure on dust facilitates wind acceleration, the actual mass-loss rate is likely determined by the rate at which gas is uplifted to the dust formation radius by convective shocks. In this view, it remains unclear if transitioning to an optically thick dust regime would significantly boost mass-loss rates. 

\paragraph{YSGs: Extrapolation to higher temperatures}

Although for many stars \gls{YSG} mass loss should have a limited impact in their evolution \citep{Zapartas_2025}, for \gls{BH} progenitors this may not be the case, since at $M_{\rm ZAMS}\,\gtrsim\,$30~$M_\odot$, \gls{CHeB}, when mass loss can considerably affect the envelope evolution, is initiated at $T_{\rm eff}\,>\,$4~kK, i.e. during the \gls{YSG} phase \AtlasIIp. Many of the mass-loss recipes for cool supergiant winds do not have any $T_{\rm eff}$ dependency at all, which may lead to several uncertainties when applied to evolutionary models, even more for binary systems, where interactions among stars can drastically affect their \gls{HR} position \citep[e.g.,][]{Dutta_2024}.

\section{Discussion: Other mass-loss uncertainties}
\label{sec:other_uncert}

The modeling of mass loss in stellar evolution is subject to several more key uncertainties, a selected list below.

\paragraph{Fe and CNO driving limits} 
A fundamental limitation in scaling mass-loss rates by total metallicity is the implicit assumption that the abundance of wind-driving elements (primarily Fe) scales linearly with the total metal content. In high-redshift environments, galaxies are often $\alpha$-enhanced, meaning the ratio of $\alpha$-elements (e.g., oxygen) to iron differs from the solar ratio \citep{Cullen_2021}. Standard mass-loss prescriptions rely on fits from atmospheric models based on Fe spectral lines that provide the necessary opacity to drive the wind. Recent hydrodynamically consistent modeling has identified a critical metallicity threshold, below which this driving mechanism fails. Below a metallicity of roughly the \gls{SMC} value of 0.2~$Z_\odot$, the Fe opacity peak becomes insufficient to drive the wind \citep{Krticka_2025}. This transition introduces significant uncertainties: while some models predict a shallow decline in mass loss \citep[$\dot{M} \propto Z^{0.42}$;][]{Vink_2021b} due to CNO efficiency, others predict a steep drop ($\dot{M} \propto Z^{1.4}$) as the wind becomes lacking of driving lines \citep{Krticka_2025}. This discrepancy, summarized in Table~\ref{tab:Z_driving_winds}, highlights a fundamental breakdown in the application of standard wind recipes. Current evolutionary codes typically extrapolate mass-loss rates to low metallicities using simple power-law scalings calibrated on Galactic (Fe-driven) winds. This approach implicitly assumes that the wind-driving mechanism remains dominated by the same set of ions throughout cosmic history. In the intermediate regime (0.1<$Z/Z_\odot$<0.2), the driving is neither purely Fe-based nor purely CNO-based; instead, $\alpha$-elements such as Silicon (Si) combined with CNO elements become the primary drivers, driving the mass-loss rates of e.g. WN stars down by roughly two orders of magnitude from the SMC levels \citep{Sander_2020a}. Extrapolating Fe-based formulations here ignores the specific influence of Si/CNO. As metallicity falls further to the CNO regime ($Z\sim0.01~Z_\odot$), the statistical forest of overlapping Fe lines disappears entirely, leaving the wind to be driven by a few discrete CNO lines. By applying a Galactic Fe-scaling to these regimes, models force a "strong wind" physics onto a "weak wind" environment. Therefore, we argue that mass-loss descriptions must move beyond simple Z-dependent scalars and adopt species-dependent driving limits.

\renewcommand{\arraystretch}{1.4}
\begin{table}[ht]
\centering
\caption{Main drivers for thin and thick winds as a function of metallicity}
\label{tab:Z_driving_winds}
\setlength{\tabcolsep}{6pt} 
\resizebox{\linewidth}{!}{
\begin{tabular}{l | c l l }
\hline
\textbf{$Z$ regime} & \textbf{$\sim Z/Z_\odot$} & \textbf{Driver} & \textbf{Key reference} \\
\hline
Galactic & 1.0\textsuperscript{a} & Fe & \cite{Abbott_1982}\\
SMC & 0.2 & Fe + CNO & \cite{Sander_2020a}\\
Fe limit & 0.1 & Si + CNO & \cite{Krticka_2025}\\
Very low \textit{Z} & 0.01 & CNO & \cite{Krticka_2009}\\
\hline
\multirow{3}{*}{Primordial}  & \multirow{3}{*}{0} & L$\alpha$ + He II & \cite{Kudritzki_2002}\\
  & & Continuum\textsuperscript{b} & \cite{Krticka_2006} \\
   & & Mechanical\textsuperscript{b} & \cite{Meynet_2006} \\
\hline
\end{tabular}
}
\parbox{\textwidth}{\footnotesize
\textsuperscript{a}{Arbitrary approximation for the Milky Way metallicity distribution.}\\
\textsuperscript{b}{Not line-driven winds}.\\
}\\
\end{table}
\renewcommand{\arraystretch}{1.0} 

\paragraph{Inflation} Envelope inflation is a rapid, non-thermal-equilibrium increase in the stellar radius, and represents a major uncertainty in massive star evolution. It is driven by near-surface layers approaching the Eddington limit, a process linked to opacity peaks from Fe or He ionization \citep{Cantiello_2009,Jermyn_2022}. While its underlying physics is not fully understood, inflation causes numerical instabilities in 1D models \citep{Agrawal_2022}. To mitigate this, ad-hoc solutions such as the \textit{use\_superad\_reduction} or \textit{MLT++} modules in {\tt MESA} are used, but this effectively alters the effective temperature by lowering temperature gradients, thereby affecting the predicted mass-loss rates and wind types. Despite this measure, inflation still occurs in some models, and the astrophysical reality of this phenomenon, along with the justification for such "stellar engineering" solutions \citep[e.g.][]{Jiang_2015}, remains debated. Regardless, the interplay between inflation and stellar winds is currently poorly understood, suggesting a need for further corrections to mass loss or the development of ad-hoc models, such as \citet{Pauli_2026}.

\paragraph{Internal mixing and rotation} Internal mixing plays an important role in the structural evolution of massive stars, even more so when rotation is included and potentially enhanced by tidal interactions with a stellar companion. This leads first to rotationally-enhanced mass-loss rates, and also to a more chemically homogeneous structure of the envelope, hence different \gls{HR} positions. \cite{Gilkis_2021} showed how rotating stars, even at $Z_\odot$, can reach sufficiently high internal mixing to keep stars chemically homogeneous and therefore making them unable to expand past the \gls{HD} limit. However, it must be highlighted that, at least at near-solar metallicity, recent models show that winds dominate the evolution of very massive stars due to their high mass loss, and internal mixing does not play a significant role \citep[e.g.,][]{Romagnolo_2024}. This is because the winds within this regime are strong enough to eject most angular momentum in the early evolutionary stages. On the other hand, there is strong observational evidence to support the concept that rotational mixing is considerably more efficient at lower metallicities \citep{Martins_2024}.

\paragraph{LBV mass loss}
In the treatment of evolved massive stars, eruptive mass loss, which can be for instance inferred by the presence of dusty \glspl{RSG} with low mass-loss rates \citep{Christodoulou_2025}, is often modeled with ad-hoc \gls{LBV} prescriptions. Some traditional \gls{LBV} formulations come from \cite{Hurley_2000}, which increases mass-loss rates by an additive factor $\dot{M}_{\rm LBV} \propto (R \sqrt{L}-1)^3(L/(6\times10^{-5})-1)$ beyond the \gls{HD} limit, or the \cite{Belczynski_2010} approach, which applies a fixed mass loss of $f_{\rm LBV} \times 10^{-4} M_{\odot} \text{yr}^{-1}$ for the same \gls{HR} regime, with $f_{\rm LBV}$ a calibrated factor usually around unity. While computationally efficient, these methods lack a dependence on the star's internal energy budget and rely on calibrations that predate modern observational and theoretical constraints. In contrast, the recent formulations by \cite{Cheng_2024} and \cite{Pauli_2026}, as well as the more canonical super-Eddington {\tt FRANEC} formulation \citep{Limongi_2018}, trigger eruptive mass loss in 1D models by identifying and ejecting envelope layers where the opacity-driven luminosity exceeds the local Eddington limit. This mechanism may naturally reproduce the absence of red supergiants above $\sim 25~M_{\odot}$ without relying on arbitrary stability thresholds, but also represents physics that overlaps with the phenomena of inflation and Fe-peak opacity driven classical \gls{WR} winds, leading to further uncertainties on which regime these phenomena dominate.

\paragraph{Mass-loss peaks and interpolation} Transitions between different mass-loss schemes in 1D stellar models, such as the onset of cool supergiant winds, the bistability jump, or the switch to optically thick winds, often create artificial jumps in mass-loss rates. These discontinuities arise because individual formulae, from different studies, are calibrated for specific stellar types and are not designed to connect smoothly, though some jumps are intentional (see e.g. the \citealt{Vink_2001} bistability jump formulation). Interpolation schemes are often adopted to smooth these transitions and avoid numerical noise from abruptly changing mass-loss rates. These computational solutions are an accepted limitation of 1D evolutionary codes, which are intended to approximate stellar evolution under an acceptable degree rather than mirror first-principle physics.

\paragraph{Mass-loss peaks and maximization} Many models do not adopt interpolation methods to transition between different mass-loss recipes, but instead parametrize mass-loss rates as the maximum value between two or more stellar winds recipes at a given timestep. For instance, many {\tt GENEC} models \citep[see e.g.][]{Eggenberger_2008,Hirschi_2025} define mass loss for WNL stars as the maximum value between \cite{Vink_2001} and \cite{Grafener_2008}. This ensures numerical stability, since usually numerical errors are more frequent with jumps in mass-loss rates or in general with more massive envelopes, and avoids the underestimation of mass loss from extrapolation. On the other hand, maximizing mass loss between two or more formulations breaks consistency with the original fits and does not necessarily represent the strength of stellar winds. The recipe that produces stronger mass loss is not necessarily the right one, and this method risks to lead to mass-loss overestimation. Whether the switch between two or more mass-loss formulations is done by linear interpolation or by taking the highest mass-loss value, stellar models lose self-consistency in favor of computational efficiency and a smoother mass-loss history.

\paragraph{Massive stars in binaries and multiples} Massive stars are not usually born as isolated entities, but with at least one stellar companion \citep{Sana_2012,Moe_2017}. On top of that, within the $M_{\rm ZAMS}$ regime of \gls{BH} progenitors it was shown by both theoretical and observational works that most of these stars reside in systems with more than two stellar components \citep{Offner_2023,Bordier_2024,deSa_2024}. Many massive stars do not therefore evolve in isolation, and their overall evolution will be affected by their stellar companion(s). Both mass transfer \citep{Renzo_2023,Landri_2024} and supernova \citep{Ogata_2021,Hirai_2023} events may considerably alter the binding energy within stellar envelopes, as well as their mass, mean molecular weight, and consequently their surface properties. This is likely to considerably impact mass loss, resulting in a non-negligible deviation from the modeled isolated evolution of the star. Furthermore, in systems hosting a neutron star, energetic pulsar winds can irradiate and ablate the companion's envelope—a process essential for shaping the evolution of spider binaries—thereby driving mass-loss rates that are entirely distinct from standard stellar winds \citep[e.g., ][]{Chen_2013,Benvenuto_2014,Misra_2025}.
All these effects represent further evidence that $\eta_{\rm trans}$ should not be a fixed factor at a given metallicity that is calculated from hypothetical chemically homogeneous conditions, but a value that can considerably vary and should be calculated following the \textit{current} surface conditions of a star (see more in Section~\ref{subsubsec:multi_scatter}). Similarly, the existence of these binary interactions imply that the $M_{\rm ZAMS}$ dependency of the mass-loss rates from \cite{Beasor_2023} and \cite{Decin_2024} limits their applicability to only single stars below the \gls{HD} limit. 

\paragraph{Magnetic fields} About 4-10\% of the observed O-type stars were shown to have strong surface magnetic fields, likely relics of prior merger events \citep{Petit_2019} which are postulated to create strong dynamo effects \citep{Wickramasinghe_2014}. The role of magnetism in mass-loss modeling remains a major source of uncertainty, as it can act as both a quenching agent and a driver of outflows.
Strong magnetic fields are thought to quench mass loss by adding an effective "gravitational" pull that keeps stellar surfaces bound \citep{Keszthelyi_2022}. However, magnetic fields can also inject mechanical energy into the atmosphere, propelling gas outward. This is particularly relevant for \glspl{YSG} and early-type \glspl{RSG}, where magnetic field reconfigurations and the generation of (magneto)acoustic or Alfvén waves can provide the sustained momentum needed to lift material off the photosphere \citep{HartmannMacGregor_1980,CharbonneauMacGregor_1995,Falceta-Goncalves_2006}. Such cool coronae or wave-driven winds may bridge the gap between purely line-driven and cool supergiant wind regimes. Furthermore, magnetic fields increase angular momentum loss via magnetic braking and fundamentally alter the internal transport of elements and angular momentum. Recent 3D simulations suggest that mergers are highly efficient at generating these stable, strong fields \citep{Vynatheya_2025}, implying that a significant fraction of massive stars may evolve under magneto-hydrodynamic conditions that current 1D prescriptions, which assume spherical symmetry and non-magnetic driving, fail to capture.

\paragraph{Pop II and Pop III}
In the primordial regime where even CNO driving is absent, the dominant mass-loss mechanism must fundamentally shift from radiation-driven to rotationally-driven or continuum-driven processes. For Population II (extremely metal-poor) and Population III (metal-free) stars, the (near-)lack of opacity implies that line-driven winds are negligible, though free-electron (Thomson) scattering may still contribute to mass loss near the Eddington limit. Consequently, rotationally-driven mechanical mass loss, which is distinct from the rotational wind mass-loss enhancement currently adopted in evolutionary models \citep[e.g., ][]{Friend_1986,Langer_1998,Maeder_2000}, may become the primary channel for mass ejection (rotation reduces the equatorial gravitational pull, which relaxes the $L/M$ ratio, and therefore increasing the Eddington factor $\Gamma_{\rm e}$). While models usually set the break-up surface angular momentum threshold around 98\% of the critical value \citep{Meynet_2006}, we argue that mass loss may be initiated considerably earlier. Standard 1D evolutionary models operate under hydrostatic equilibrium that explicitly neglects the radial inertial term in the momentum equation. By defining the critical limit purely through static force balance rather than a hydrodynamic launch condition, these models essentially visualize the stellar surface as "static" even at breakup. This approximation ignores the dynamic pressure of the outflow and the complex effects of turbulence \citep[e.g.,][]{Kee_2021}, which can establish a non-zero velocity baseline and effectively lower the angular momentum threshold for mass shedding. Consequently, the implementation of mechanical mass loss in 1D simulations often represents a numerical compromise rather than a resolved physical process. Codes typically employ algorithmic "patches", such as implicit mass-loss boosts \citep[][]{Ekstrom_2008}, that instantaneously remove mass and angular momentum to force the star to remain sub-critical. This zero-velocity visualization fails to capture the anisotropy of the outflow, where mass loss likely manifests differently at the pole and the equator, as well as potentially underestimating the terminal velocity of stellar material from the surface.

\paragraph{Pulsations}
Pulsational instabilities are not usually included in standard mass-loss formulations for massive stars. While 1D stellar evolution codes like {\tt MESA} can be coupled with linear pulsation modules such as {\tt GYRE} \citep{Townsend_2013,Townsend_2018,Goldstein_2020,Sun_2023} to identify unstable modes, these tools are limited to linear stability analysis. They cannot self-consistently predict the non-linear saturation amplitudes required to determine mass fluxes. Consequently, accurately resolving pulsation-driven mass loss requires full 3D radiation-hydrodynamics simulations to capture the associated shocks and turbulence \citep{Jiang_2015,Freytag_2017,Jiang_2018}, or the use of 1D modules that approximate these effects, such as turbulent pressure models \citep{Fuller_2024} or phenomenological prescriptions like those of \cite{Yoon_2010}. However, more than a decade of development in evolutionary models has introduced critical additions to better model convection beyond the standard mixing length theory approach. These computational improvements have enabled highly challenging simulations of \gls{RSG} envelopes, helping to reconcile the physics of mass loss with supernova observations. Building on the work of \citep{Yoon_2010}, these updated pulsation-driven winds and their consequent shock-aided ejections now provide a self-consistent way to explain the observed features of Type II supernovae, a finding recently and independently established by several groups \citep[e.g.][]{Bronner_2025,Suzuki_2025,Laplace_2026,Sengupta_2026}.

\paragraph{Massive stars still accrete during \gls{MS}}
Massive stars are a testbed for stellar wind theories because of their strong mass-loss rates. However, some of these stars may continue to accrete material during their \gls{MS}, as they remain embedded within their natal clouds until their feedback effectively clears the surrounding gas \citep{PallaStahler_1993}. The time at which a star begins its wind-driven mass loss may therefore not coincide with its \gls{ZAMS}.
For a given accretion rate and initial seed, the accretion timescale of a protostar is directly proportional to its current mass \citep{Wainer_2026}. The degree to which massive star formation can be described as a "scaled-up" version of low-mass star formation remains currently under debate \citep[see][for a recent review]{Beuther_2025}. Additionally, very massive star formation imposes further challenges \citep{Krumholz_2015}, since the accretion timescale may exceed the free-fall time for the cloud, therefore extending the accretion to the \gls{MS} itself \citep{Grudic_2022}. \cite{Wainer_2026} has shown that accretion can continue up to 0.5-2~Myr after \gls{ZAMS}. This does not represent a critical factor in the mass-loss history of the lower-mass massive stars (despite neither a negligible one), but for very massive stars this may represent a considerable fraction of their nuclear timescales, since their total lifetime is around 3 to 5~Myr. If very massive stars were accreting for that long, this may completely erase wind-driven mass loss for at least the first 10\% of their \gls{MS}, further increasing the uncertainties on the coupling of rotation and internal mixing with mass-loss rates.

\section{Conclusions}

In this work 
we have presented a systematic comparative analysis of the stellar wind prescriptions widely used in modern stellar evolution codes for massive stars. 
By examining the three primary mass-loss regimes (optically thin line-driven winds, optically thick Wolf-Rayet winds, and cool supergiant winds) we have highlighted the deep and often contradictory assumptions embedded in current models. The choice of a specific mass-loss recipe is not a minor implementation detail but a dominant source of uncertainty that fundamentally shapes the predicted evolution of massive stars.

Our key findings reveal significant, and in some cases irreconcilable, discrepancies across all investigated mass-loss regimes:

\begin{itemize}
    \item Optically thin winds: We find that the predicted mass-loss rates can differ by more than an order of magnitude depending on the chosen prescription. The treatment of wind clumping and the existence, strength, and temperature location of the bistability jump(s) remain major points of contention, leading to vastly different mass-loss histories even for identical stars.
    \item Optically thick winds: We identified critical flaws in the criteria used to trigger the formation of a Wolf-Rayet star. We show that the commonly used Eddington factor $\Gamma_{\rm e}$ may act as an inconsistent proxy for the initiation of optically thick winds, if applied outside its domain of validity. Furthermore, we demonstrate that the alternative multi-scattering $\eta$ criterion for wind efficiency is highly model-dependent, and that its current implementation in evolutionary codes sacrifices consistency with a star's instantaneous surface properties to artificially anchor the model to the input requirements of the \cite{Vink_2011} parameterization. To address this, we provide recommendations for improving the multi-scattering criterion and introduce a new, model-agnostic framework for this transition. By calculating wind efficiency and escape velocities based on the star's current physical state rather than hypothetical chemically homogeneous conditions, this flexible approach ensures greater physical consistency across the mass spectrum.
    \item Cool supergiant winds: The landscape of available prescriptions is equally divergent. Recent models based on new observational constraints predict mass-loss rates that can be orders of magnitude lower than the canonical prescriptions traditionally used in stellar evolution, but with similar or even higher mass loss than legacy models at $\log (L/L_\odot)$\,>\,5.5, i.e. the extrapolation regime for these winds past the \acrfull{HD} limit. 
\end{itemize}

We also introduce the "cool Wolf-Rayet problem", a regime for which no dedicated, physically consistent model currently exists. Consequently, models must employ ad-hoc workarounds. For instance, some models extrapolate weaker optically thin winds into this cooler domain \citep{Romagnolo_2024,GormazMatamala_2025}
, which is more justified by recent theoretical studies \citep{BerniniPeron_2025,Lefever_2025} while others adopt stronger, optically thick wind prescriptions calibrated from warmer Wolf-Rayet stars \citep{Glebbeek_2009,Sabhahit_2023}
, which instead better respect the Humphreys-Davison limit \citep{Boco_2025}
. The uncertainty extends to the transition criteria themselves. The use of $\Gamma_{\rm e}$-dependent physics is formally valid only for fully ionized plasmas where electron scattering dominates, a condition met at effective temperatures $T_{\rm eff}\,\geq\,$30~kK, with a potential approximated extension down to 15~kK due to the still relatively strong contribution of free-electron scattering. Below this threshold, the recombination of hydrogen and helium contributes considerably to the opacity, and the physical basis for the classical $\Gamma_{\rm e}$ calculation collapses. Applying transition criteria and mass-loss rates based on the classical $\Gamma_{\rm e}$ in this regime, despite being partially justified because it still acts as a proxy for the $L\,/\,M$ ratio, is a fundamental flaw that forces models to rely on uncertain extrapolations rather than a self-consistent physical framework.

\section*{Software acknowledgements}
{\small
This work used the following software packages: \texttt{matplotlib} \citep{Hunter:2007}, \texttt{numpy} \citep{numpy}, and \texttt{pasta-marker} \citep{Pasta_2024,Pasta_2025}.\\
The {\tt MESA} models used are available at \url{github.com/AmedeoRom/Stellar_Winds_Atlas}

\vspace*{-0.2cm} 

\section*{Acknowledgements}
AR acknowledges financial support from the European Research Council for the ERC Consolidator grant DEMOBLACK, under contract no. 770017 and from the German Excellence Strategy via the Heidelberg Cluster of Excellence (EXC 2181 - 390900948) STRUCTURES. ACGM thanks the support from project 10108195 MERIT (MSCA-COFUND Horizon Europe). Computations for this article have been performed using the computer cluster at CAMK PAN.
The authors thank Lucas M. de Sá, Daniel Pauli, Cristiano Ugolini, Jan Eldridge, Mathieu Renzo, Gemma Gonzalez i Tora, JD Merritt, Lumen Boco, Gautham Sabhahit, Michela Mapelli, Emmanouil Zapartas, Andreas Sander, Earl Bellinger, Harim Jin, Radek Smolec, Sylvia Ekström, and Jiří Krtička for their feedback and the constructive discussions.

\bibliographystyle{aasjournal}

\bibliography{oja_template}








 

\end{document}